# *Ab Initio* Prediction of Excited State and Polaron Effects in Transient XUV Measurements of α-$Fe_2O_3$

Isabel M. Klein,[1] Hanzhe Liu,[1] Danika Nimlos,[1] Alex Krotz,[1] Scott K. Cushing[1,†]

[1]Division of Chemistry and Chemical Engineering, California Institute of Technology, Pasadena, CA 91125, USA.

[†]Corresponding author. Email: scushing@caltech.edu

Abstract

Transient X-ray and extreme ultraviolet (XUV) spectroscopies have become invaluable tools for studying photoexcited dynamics due to their sensitivity to carrier occupations and local chemical or structural changes. One of the most studied materials using transient XUV spectroscopy is α-$Fe_2O_3$ because of its rich photoexcited dynamics, including small polaron formation. The interpretation of carrier and polaron effects in α-$Fe_2O_3$ is currently done using a semi-empirical method that is not transferrable to most materials. Here, an ab initio, Bethe-Salpeter equation (BSE) approach is developed that can incorporate photoexcited state effects for arbitrary materials systems. The accuracy of this approach is proven by calculating the XUV absorption spectra for the ground, photoexcited, and polaron states of α-$Fe_2O_3$. Furthermore, the theoretical approach allows for the projection of the core-valence excitons and different components of the X-ray transition Hamiltonian onto the band structure, providing new insights into old measurements. From this information, a physical intuition about the origins and nature of the transient XUV spectra can be built. A route to extracting electron and hole energies is even shown possible for highly angular momentum split XUV peaks. This method is easily generalized to K, L, M, and N edges to provide a general approach for analyzing transient X-ray absorption or reflection data.

## Introduction

Transient extreme ultraviolet (XUV) spectroscopy has been used to measure electron and hole populations, phonon modes, polaronic states, and charge-transfer in layered photoelectrodes – making it a new favorite tool for studying photocatalytic and photoelectrochemical systems.[1–9] The low energy (10-150 eV) and long wavelength of the XUV transition makes it more sensitive to delocalized valence states relative to soft or hard X-rays. Local structural changes can be indirectly probed, including polarons and acoustic phonons. However, the combination of this information makes the interpretation of photoexcited XUV spectra particularly challenging. Unlike molecules where orbitals are localized, solids are subject to the added complexity of cluttered band structures, core-hole screening, and other many-body effects that must be taken into account when analyzing XUV spectra.[10–15] Hard X-ray transitions are computed using a modified pseudopotential in a supercell to replicate core-hole effects, however this method is not accurate for modeling low energy XUV transitions.[16]

Hematite (α-$Fe_2O_3$) has become the prototypical material for transient XUV experiments because of its many photoexcited dynamics, including small polaron physics, and relevance to photocatalysis and photoelectrochemistry.[13,17–25] Small polaron formation traps photoexcited electrons at Fe sites, significantly limiting carrier mobilities and lifetimes.[13,26–28] Transient XUV spectroscopy has been used to evaluate how defects, nanostructuring, doping, surface decoration, and other material aspects can modulate small polaron formation to improve performance. [2,13,17–20,29] To date, these measurements have all used the assumptions of a semi-empirical atomic multiplet theoretical approach, a change in oxidation state upon excitation, and polaron splitting of the 3p core level.[4,13,17,19,20,25,30] These approximations are accurate for $\alpha$-$Fe_2O_3$ because of the highly localized Fe 3d orbitals, however, this approach is not generalizable, with new models generally developed for each new material studied.[1,3,8,31]

Here, we develop an ab initio Bethe-Salpeter equation (BSE) approach that can accurately predict photoexcited changes in the transient XUV spectra, including photoexcited electron and hole distributions, polarons, and thermal effects. The theory is used to reproduce previous results for $\alpha$-$Fe_2O_3$, both to test its accuracy and to show that it can provide new insights into the measured dynamics from past experimental data. Our modified BSE method allows for the peak structure of the Fe $M_{2,3}$ edge to be mapped directly onto the band structure. The spectrum can be further decomposed into the components of the X-ray interaction Hamiltonian to help develop a physical intuition into the origins and nature of the ground and photoexcited spectra. Going beyond previous models, we show that electron and hole energies could theoretically be extracted from an angular momentum split XUV spectrum, which was previously only possible in highly screened semiconductors.[3,8] The underlying theoretical approach is easily extended to any K, L, M, or N edge as long as an accurate underlying DFT or GW calculation is possible for the material.[32–38] The methods in this paper provide a general and facile technique for understanding time-resolved X-ray spectra for synchrotron, table-top XUV, or free electron laser measurements. Given the complexity of these experiments, it is invaluable to be able to predict whether the excited state phenomena of interest will be measurable for a given X-ray edge.

## Methods

The theoretical approach is based on modifications to the Obtaining Core Excitations from *Ab initio* electronic structure and the NIST BSE solver (OCEAN) code.[39,40] The OCEAN code has previously been verified for ground state K, L, M, and N edge calculations.[38,41,42] The base code is expanded here in three ways. First, the core-valence excitons that underly the XUV absorption spectrum are calculated and can be projected onto the band structure. Second, the X-ray transition matrix elements are decomposed into their fundamental components and can also be projected onto the band structure. Finally, photoexcited changes to the band occupations, phonon modes, and polaronic-type effects are included for the prediction of excited state spectra.

The ground state core-level spectrum is calculated in three steps. The core-level transition matrix elements are first determined by projecting an all-electron atomic calculation of the core levels to a density functional theory (DFT+U) calculation of the valence wavefunctions using a projector augmented wave (PAW)-style optimal projector functions (OPFs).[39,40,43] Second, screening of the core electrons in the presence of the core-hole is determined self-consistently.[40,43] Atomic multiplet effects are included through core-hole spin-orbit splitting and atomic multiplet interactions.[39] Finally, the ground state absorption spectrum is obtained by iteratively solving the BSE using a Haydock recursive algorithm.[40,43] Alternatively, the real-space wavefunction of the core-valence excitons can be calculated using a Generalized Minimal Residual (GMRES) method.

The calculated core-valence excitons are projected onto the band structure using an overlap integral between the GMRES calculated real-space wavefunction and the energy and momentum specific wavefunction from the ground state DFT calculation. This integral maps the calculated XUV spectrum onto the ground state band structure. This projection can be decomposed into the contributions from separable terms of the X-ray transition Hamiltonian, including the spin-orbit angular momentum coupling, core-hole screening, atomic multiplet effects, and higher-order BSE terms. This calculation can be repeated for all the modeled photoexcited states. These calculations identify the dominant core-valence interactions that lead to the XUV spectrum and can provide some physical intuition into the origins and nature of experimentally observed photoexcited changes.

Excited state effects are included under the adiabatic approximation that the core-hole lifetime is shorter than any modeled optically excited state processes. Since the excited state predictions of interest in this study deal with electron-electron and electron-phonon scatterings that occur on the few-femtosecond or longer timescales, this assumption is valid, however it would not hold true for few cycle or attosecond effects.[39] For the excited state calculations, the valence and conduction band occupations are modified to reflect the change in carrier occupation due to photoexcitation (Figure S25 and S26). The polaron is modeled by first replacing one of the iron sites with a cobalt atom to introduce an additional localized electron, relaxing the system to determine the local lattice change, and then replacing the original iron atom for the final self-consistent calculation.[44–46] Within this approximation, the polaron expands the Fe-O bond length by 0.4 % and contracts the Fe-Fe bond by 1.8 %, consistent with other reports.[44,47] The resultant wavefunctions and lattice configurations are used as the input for calculating the XUV absorption using the OCEAN code (Figure S2). Given the differential measurement type calculated in this study, only the locally effected photoexcited atoms are included in the polaron model, as the background spectrum is subtracted out. We further distinguish the polaron spectral features from thermal isotropic lattice expansion. To model thermal expansion, the undistorted $\alpha$-$Fe_2O_3$ unit cell was isotopically expanded up to 0.8 %, corresponding to a temperature increase from 300 K to 350 K, and the OCEAN calculation was run with the expanded lattice (Figure S3).[48] In all cases, the differential transient XUV spectra are calculated, subtracting the calculated ground state spectrum from the relevant excited state spectrum. Full details of ground state and excited state calculations are provided in the SI.

## Results and Discussion

Figure 1a shows schematically the lattice change that occurs during small polaron formation in $\alpha$-$Fe_2O_3$. Calculated values for the lattice distortions are provided in the SI. For the Fe $M_{2,3}$ edge of $\alpha$-$Fe_2O_3$, the XUV absorption corresponds to the transition between the Fe 3p core level and the unoccupied Fe 3d DOS, as modified by interactions with the created Fe 3p core-hole. Figure 1b shows the calculated, broadened ground state absorption spectrum of the Fe $M_{2,3}$ edge (Figure 1b, blue trace) as compared to an experimental measurement (Figure 1b, black trace). The OCEAN calculation (Figure 1b, orange trace) is broadened with an energy-dependent Gaussian with a high energy Fano correction for direct comparison with previous reports.[3,38] The broadening method adequately accounts for the different lifetimes of the angular momentum and atomic multiplet split peaks, as well as the Fano-type line shape of the $M_{2,3}$ edge, and is fully discussed in the SI.

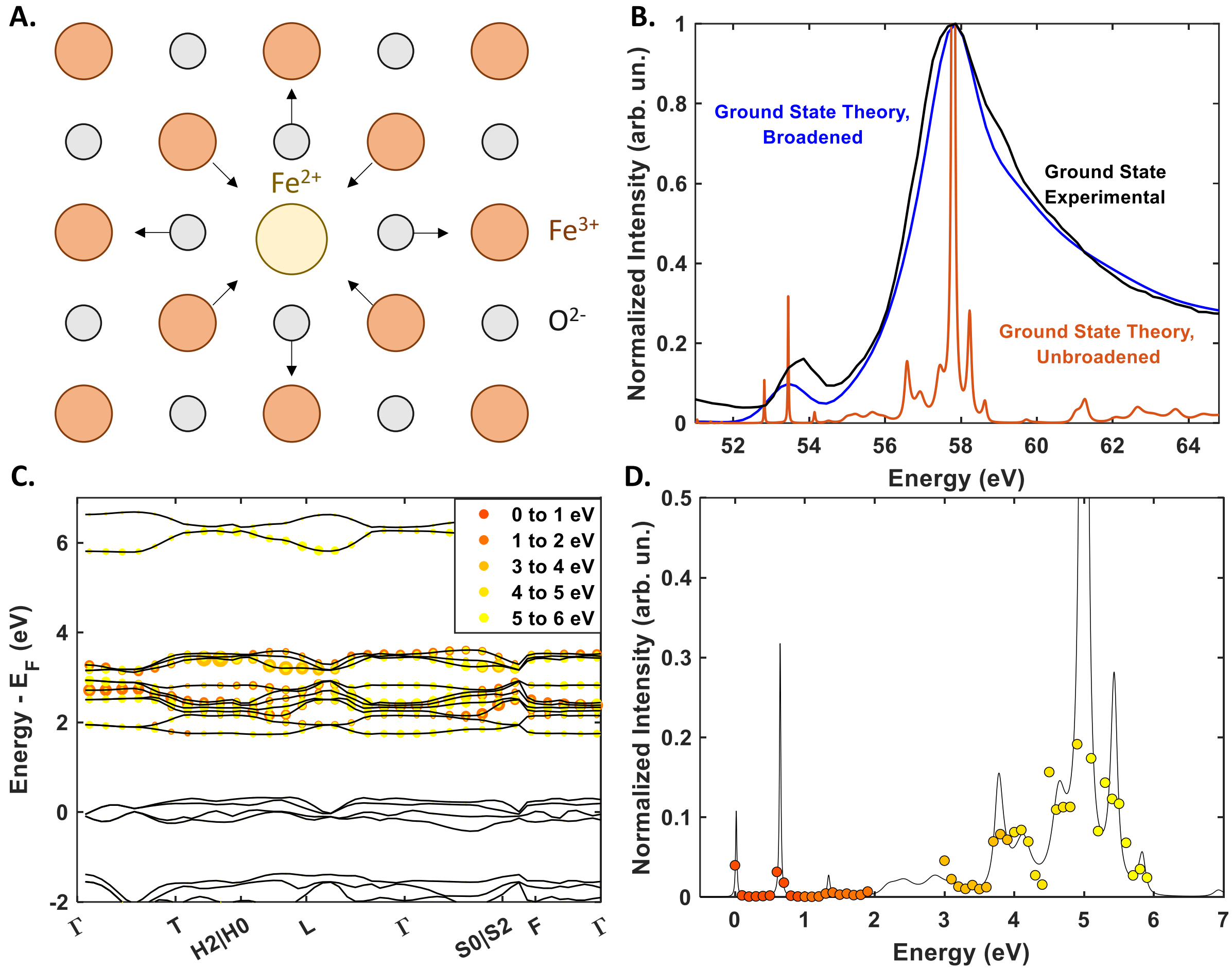

Figure 1. Simulation of the Fe $M_{2,3}$ edge of α-$Fe_2O_3$. (a) Schematic showing how the initial excitation of α-$Fe_2O_3$ leads to an oxygen (grey) to iron (orange) charge-transfer transition. Electron-phonon scattering then leads to charge localization as a small polaron, locally modifying the lattice. (b) The experimental (black), OCEAN calculated unbroadened (orange), and OCEAN calculated with energy-dependent broadening (blue) spectra are compared. (c) Distribution of the core-valence exciton density in the band structure. (d) Corresponding distribution of the core-valence exciton density over the unbroadened XUV absorption spectrum (solid black line).

To a first approximation, the XUV absorption spectrum should match the dipole-allowed transitions to the unoccupied Fe 3d density of states (DOS) (Figure S4). This approximation is true if the core-hole doesn't perturb to the final transition state. However, the core-hole strongly perturbs the final transition state through a variety of core-hole screening, angular momentum, and atomic multiplet effects.[10] The core-valence exciton density for the α-$Fe_2O_3$ XUV absorption spectrum (Figure 1d) is projected onto the band structure (Figure 1c), where the size of the dots refers to the amplitude and the color denotes the energy range within the spectrum. Figure 1c reveals how strongly the core-hole perturbation splits the originally narrow Fe 3d conduction band into absorption peaks more than 4 eV apart, highlighting the difficulty in modeling and interpreting the XUV spectra.[18]

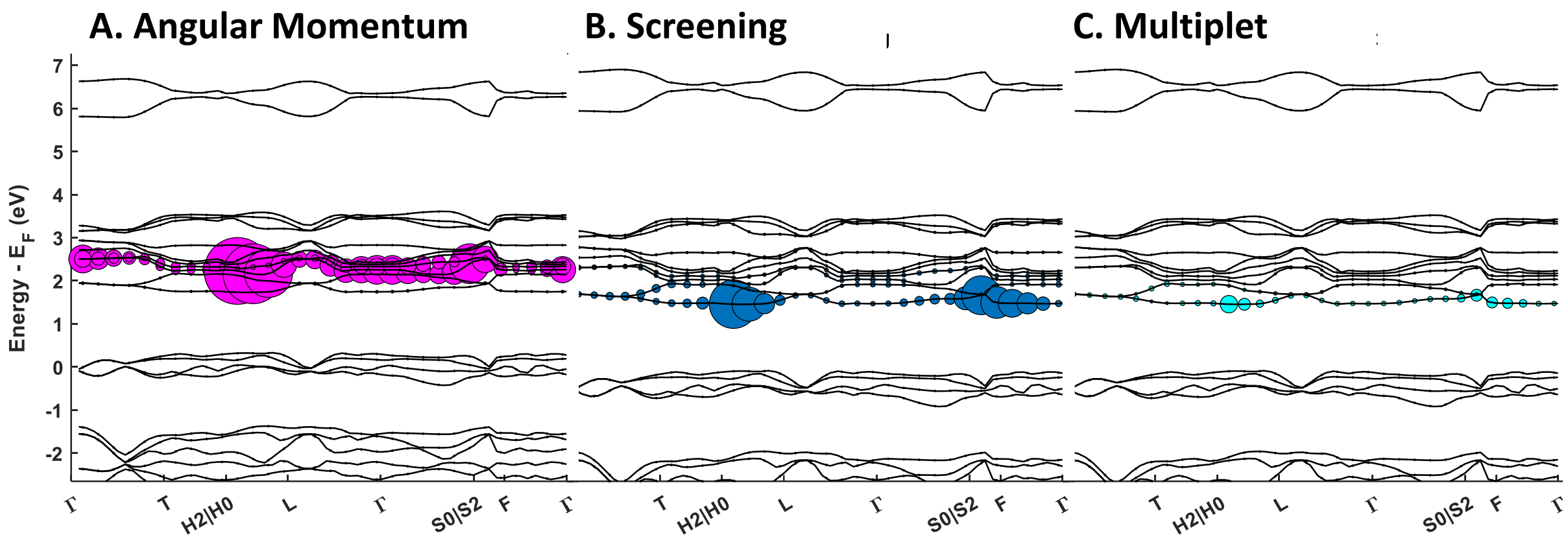

Figure 2. Decomposition of the Core-Valence Exciton into X-ray Transition Hamiltonian Components. The (a) spin-orbit angular momentum coupling, (b) core-hole screening, and (c) atomic multiplet components of the core-valence exciton of the ground state are projected and normalized onto the band structure. The different Hamiltonian components determine how the valence DOS is distorted into the XUV absorption spectrum. The bubble, ladder, and Slater components of the XUV transition Hamiltonian are shown in Fig S13.

To further understand the relationship between the band structure and the measured XUV spectrum, the core-valence exciton density can be broken down into the dominant terms of the X-ray transition Hamiltonian. Figure 2 shows that the XUV absorption spectrum comes from a combination, in decreasing order of influence, of spin-orbit angular momentum coupling, core-hole screening, and atomic multiplet effects between the $3p^6 3d^5$ ground state and the $3p^5 3d^6$ XUV excited state.[15] The angular momentum contribution refers to the spin-orbit coupling of the core-hole and the valence state, the screening term describes the ability of the valence electrons to screen the core-valence exciton, and multiplet effects arise from the overlap between the core and valence wavefunctions in the final $3p^5 3d^6$ state. The angular momentum contribution is mainly in the middle of the Fe 3d bands and explains the splitting into two dominant peaks in the XUV absorption spectrum (Figure 1). The screening contribution is less pronounced and concentrated near the bottom of the conduction band, mainly between the $\Gamma$ and $L$ points. The multiplet effects are minimal, matching the many-body bubble and ladder terms (Figure S13).

Photoexcitation of $\alpha$-$Fe_2O_3$ initiates a ligand-to-metal charge transfer between the majority O 2p valence band and the majority Fe 3d conduction bands (Figures S5 and S7). Photoexcitation changes the occupation, or state-filling, of the conduction band. Since the Fe 3d orbitals that make up the conduction band are localized, this process is often referred to as an effective reduction of the $Fe^{3+}$ center to a $Fe^{2+}$. However, it is more accurately described as a change in state-filling, which refers to the photoexcited addition of electrons to the conduction band and the presence of holes in the valence band. According to the X-ray Hamiltonian components shown in Figure 2, photoexcited electrons will decrease angular momentum splitting effects, while increasing the screening of the core-valence excitons. These changes will immediately red shift the overall XUV absorption spectrum relative to the ground state. In comparison, if small polarons and thermal expansion are present, there will be a change in wavefunction overlap that will mostly affect the angular momentum contributions, leading to a change in peak splitting and ordering.

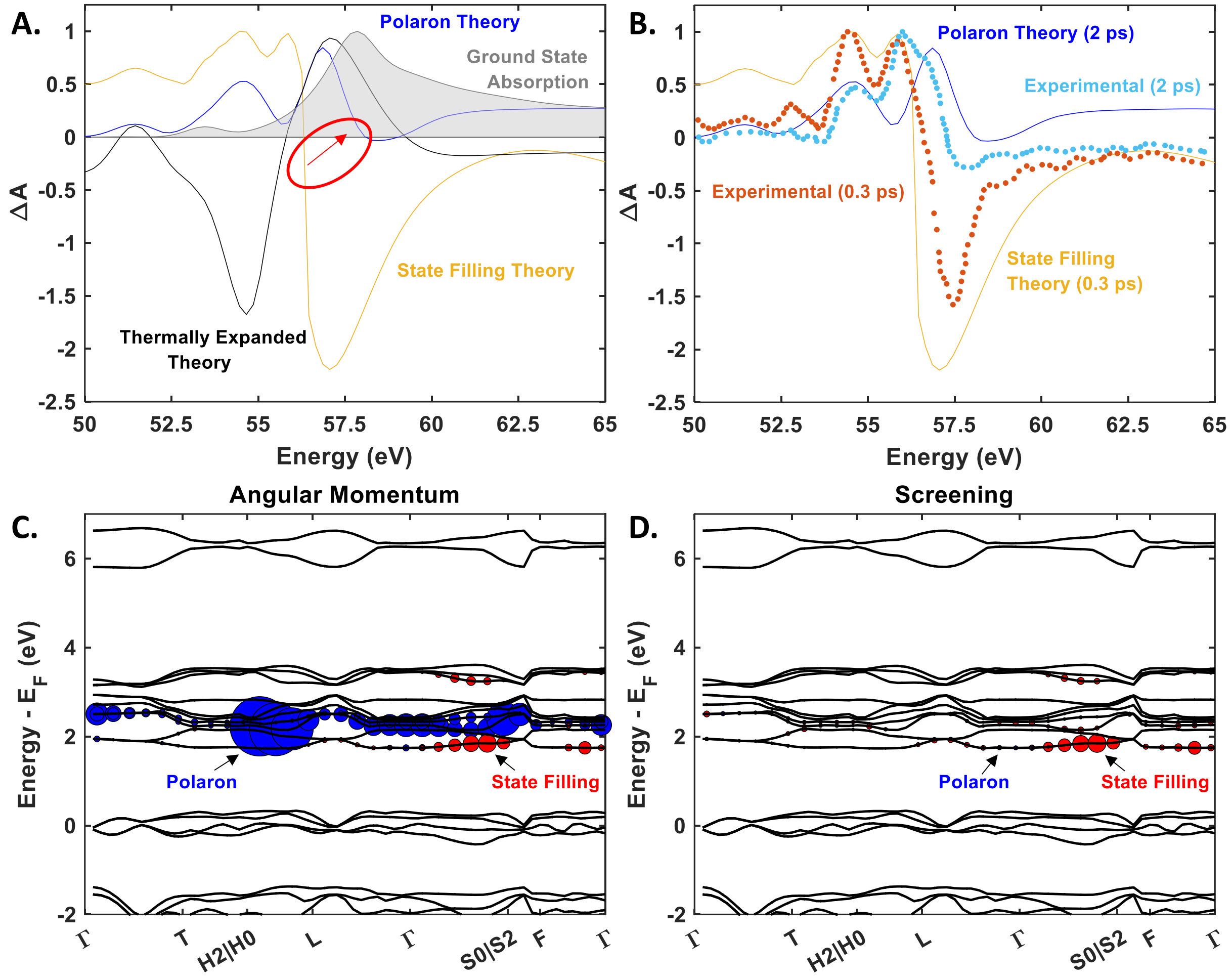

Figure 3. Excited state calculation of the Fe $M_{2,3}$ edge of $\alpha$-$Fe_2O_3$. (a) Theoretically predicted differential absorption between the ground state and the photoexcited (yellow), polaron (blue), and thermally expanded (black) states. The red circle and arrow indicate the shift of the zero crossing commonly used to measure polaron formation rates. The light grey curve is the ground state XUV absorption of $\alpha$-$Fe_2O_3$ to provide a reference of where in the spectrum the differential features arise. (b) Calculated differential absorption compared to experimental data at two time points. The calculated state-filling (yellow) and polaron state (dark blue) are shown as lines while the experimentally measured excited state at 0.3 ps (orange dots) and at 2 ps (light blue dots) from ref. 12 are shown as dots. (c) Differential plots of the change in angular momentum contributions for the calculated state-filling (red) and polaron (blue) excited states. (d) Differential plots of the change in screening contributions for state-filling (red) and polaron (blue) excited states.

Figure 3a compares the calculated differential spectra for the photoexcited, polaron, and thermally expanded states. The ground state absorption spectrum is also shown for reference of where in the spectrum differential features arise. Figure 3b compares the calculated differential spectra to measured experimental data from reference 12 at 0.3 ps and 2 ps, respectively, to show the accuracy of the theoretical approach.[13] The thermal differential signature is not consistent with experimental data, in particular the trace at 2 ps, further confirming the assignment of these measured dynamics to the polaron state. Including photoexcited carriers in the polaron calculation increases the magnitude of the negative feature above 57.5 eV, as shown in Figure S24. The experimental data lies in between the polaron models with and without photoexcited carriers, as in the experiment the bands are not filled, but fractional occupations are unstable using the current theoretical approach.

Figure 3c and 3d show the change in the angular momentum and screening contributions to the X-ray transition Hamiltonian in the photoexcited state-filling and polaron cases as compared to the ground state. Consistent with the intuitive understanding from Figure 2, Figure 3c and 3d show that state-filling (red) changes both screening and angular momentum components of the core-valence exciton with the same relative magnitude. The polaron state (blue), meanwhile, almost exclusively changes the angular momentum coupling. Since state-filling better screens the core-hole perturbation the resultant spectrum is red-shifted, leading to the derivative-like features seen in Figure 3a. The change in the angular momentum coupling in the polaron state leads to a mostly positive feature that resembles the two angular momentum split peaks getting closer together in energy. The progression from a differential-like feature to a mostly positive line shape creates the signature blue-shift of the zero-crossing point (Figure 3a) that is used as a reference for polaron formation times in literature.[13,19,20,24]

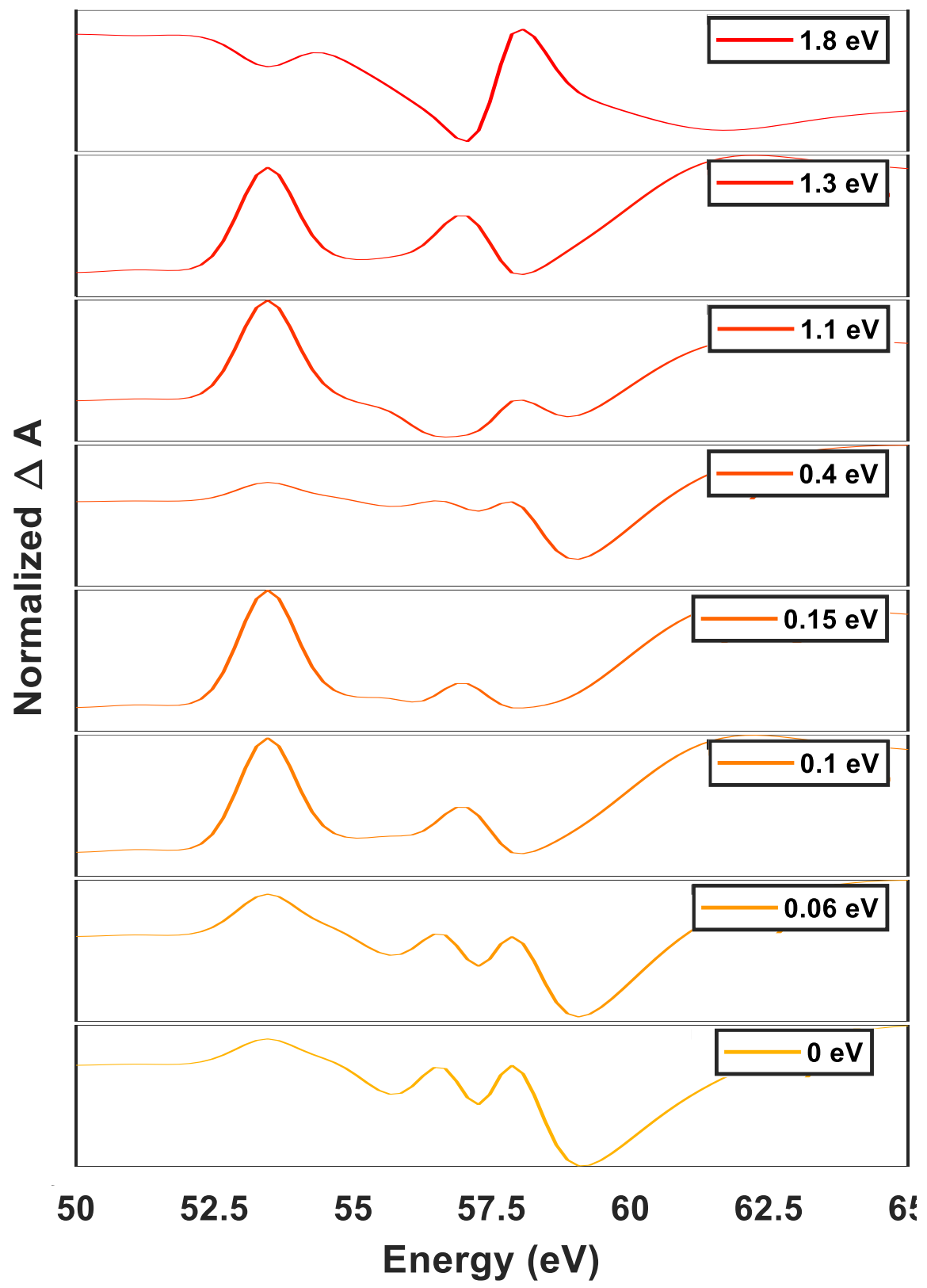

Figure 4. Excitation Energy Dependent Calculations. Differential spectra around the Fe $M_{2,3}$ edge of a-$Fe_2O_3$ following state-filling with different excitation energies above the band gap.

The Hamiltonian components in Figures 2 and 3 are not linearly distributed throughout the conduction band. As photoexcited carriers thermalize, the differential spectra will therefore shift and change in a nonlinear fashion, making extraction of electron and hole energies difficult. Figure 4 presents the differential spectra calculated for electrons occupying several energies above the conduction band minimum. Angular momentum coupling is most changed when carriers are in the middle of the conduction band (Figure 3c), which is best represented by the rise and growth of the positive feature around 53 eV. Screening is increasingly changed as carriers approach the conduction band minimum (Figure 3d) as indicated by the differential negative feature at 59 eV increasing in depth as the spectrum is further red-shifted (especially in the last ~0.2 eV above the conduction band minimum). The nonlinear evolution of Figure 4 explain why the extraction of hot carrier distributions was believed to be impossible in the presence of strong angular momentum coupling, especially when compared to highly-screened materials like Si or Ge where the X-ray structure resembles the unoccupied density of states.[3,8] While complex, the computational approach developed here does prove that hot carriers and holes can be monitored by their unique spectral signatures, it just requires a back-extraction algorithm like we have used for ZnTe.[49] For $\alpha$-$Fe_2O_3$, the polaron state forms on the timescale of the first electron phonon scattering and dominates the differential absorption, preventing such a procedure.

## Conclusions and Outlook

An ab initio method is developed to model excited state effects in the transient XUV spectra of solid-state materials. The method is verified by accurately predicting the complex photophysics of $\alpha$-$Fe_2O_3$ and small polaron formation, one of the most studied materials using transient XUV spectroscopy. The ab initio method allows for an in-depth analysis of the XUV spectrum by projecting the core-valence exciton density and major X-ray transition Hamiltonian components on to a material's band structure. The decomposition of the XUV spectrum facilitates an intuitive understanding into the origin of excited state changes, even in a highly angular-momentum-splitting controlled XUV absorption spectrum. If polaron effects weren't dominant, it could even be possible to extract electron and hole energies from the transient XUV spectra, using the theoretical approach outline here. The technique has obvious extension to X-ray signatures of intermediate and large polarons in other photocatalytic materials, such as those in $BiVO_4$, $TiO_2$, ZnO, and various perovskites. More broadly, however, the technique presents a BSE centered, materials-independent method to interpret transient K, L, M, and N edge measurements, whether made using synchrotrons, table-top XUV spectrometers, or X-ray free electron lasers.

References

(1) Cushing, S. K.; Porter, I. J.; Roulet, B. R. de; Lee, A.; Marsh, B. M.; Szoke, S.; Vaida, M. E.; Leone, S. R. Layer-Resolved Ultrafast Extreme Ultraviolet Measurement of Hole Transport in a Ni-TiO2-Si Photoanode. *Science Advances* **2020**, *6* (14), eaay6650. https://doi.org/10.1126/sciadv.aay6650.

(2) Biswas, S.; Husek, J.; Londo, S.; Fugate, E. A.; Baker, L. R. Identifying the Acceptor State in NiO Hole Collection Layers: Direct Observation of Exciton Dissociation and Interfacial Hole Transfer across a $Fe_2O_3$/NiO Heterojunction. *Phys. Chem. Chem. Phys.* **2018**, *20* (38), 24545–24552. https://doi.org/10.1039/C8CP04502J.

(3) Cushing, S. K.; Zürch, M.; Kraus, P. M.; Carneiro, L. M.; Lee, A.; Chang, H.-T.; Kaplan, C. J.; Leone, S. R. Hot Phonon and Carrier Relaxation in Si(100) Determined by Transient Extreme Ultraviolet Spectroscopy. *Structural Dynamics* **2018**, *5* (5), 054302. https://doi.org/10.1063/1.5038015.

(4) Biswas, S.; Husek, J.; Baker, L. R. Elucidating Ultrafast Electron Dynamics at Surfaces Using Extreme Ultraviolet (XUV) Reflection–Absorption Spectroscopy. *Chem. Commun.* **2018**, *54* (34), 4216–4230. https://doi.org/10.1039/C8CC01745J.

(5) Zürch, M.; Chang, H.-T.; Borja, L. J.; Kraus, P. M.; Cushing, S. K.; Gandman, A.; Kaplan, C. J.; Oh, M. H.; Prell, J. S.; Prendergast, D.; Pemmaraju, C. D.; Neumark, D. M.; Leone, S. R. Direct and Simultaneous Observation of Ultrafast Electron and Hole Dynamics in Germanium. *Nature Communications* **2017**, *8* (1), 1–11. https://doi.org/10.1038/ncomms15734.

(6) Zürch, M.; Chang, H.-T.; Kraus, P. M.; Cushing, S. K.; Borja, L. J.; Gandman, A.; Kaplan, C. J.; Oh, M. H.; Prell, J. S.; Prendergast, D.; Pemmaraju, C. D.; Neumark, D. M.; Leone, S. R. Ultrafast Carrier Thermalization and Trapping in Silicon-Germanium Alloy Probed by Extreme Ultraviolet Transient Absorption Spectroscopy. *Structural Dynamics* **2017**, *4* (4), 044029. https://doi.org/10.1063/1.4985056.

(7) Lin, M.-F.; Verkamp, M. A.; Leveillee, J.; Ryland, E. S.; Benke, K.; Zhang, K.; Weninger, C.; Shen, X.; Li, R.; Fritz, D.; Bergmann, U.; Wang, X.; Schleife, A.; Vura-Weis, J. Carrier-Specific Femtosecond XUV Transient Absorption of $PbI_2$ Reveals Ultrafast Nonradiative Recombination. *J. Phys. Chem. C* **2017**, *121* (50), 27886–27893. https://doi.org/10.1021/acs.jpcc.7b11147.

(8) Kaplan, C. J.; Kraus, P. M.; Ross, A. D.; Zürch, M.; Cushing, S. K.; Jager, M. F.; Chang, H.-T.; Gullikson, E. M.; Neumark, D. M.; Leone, S. R. Femtosecond Tracking of Carrier Relaxation in Germanium with Extreme Ultraviolet Transient Reflectivity. *Phys. Rev. B* **2018**, *97* (20), 205202. https://doi.org/10.1103/PhysRevB.97.205202.

(9) Cirri, A.; Husek, J.; Biswas, S.; Baker, L. R. Achieving Surface Sensitivity in Ultrafast XUV Spectroscopy: M2,3-Edge Reflection-Absorption of Transition Metal Oxides. 25.

(10) Liu, H.; Klein, I. M.; Michelsen, J. M.; Cushing, S. K. Element-Specific Electronic and Structural Dynamics Using Transient XUV and Soft X-Ray Spectroscopy. *Chem* **2021**, *7* (10), 2569–2584. https://doi.org/10.1016/j.chempr.2021.09.005.

(11) Zhang, K.; Ash, R.; Girolami, G. S.; Vura-Weis, J. Tracking the Metal-Centered Triplet in Photoinduced Spin Crossover of Fe(Phen)32+ with Tabletop Femtosecond M-Edge X-Ray Absorption Near-Edge Structure Spectroscopy. *J. Am. Chem. Soc.* **2019**, *141* (43), 17180–17188. https://doi.org/10.1021/jacs.9b07332.

(12) Ryland, E. S.; Zhang, K.; Vura-Weis, J. Sub-100 Fs Intersystem Crossing to a Metal-Centered Triplet in Ni(II)OEP Observed with M-Edge XANES. *J. Phys. Chem. A* **2019**, *123* (25), 5214–5222. https://doi.org/10.1021/acs.jpca.9b03376.

(13) Carneiro, L. M.; Cushing, S. K.; Liu, C.; Su, Y.; Yang, P.; Alivisatos, A. P.; Leone, S. R. Excitation-Wavelength-Dependent Small Polaron Trapping of Photoexcited Carriers in α-Fe2O3. *Nature Mater* **2017**, *16* (8), 819–825. https://doi.org/10.1038/nmat4936.

(14) de Groot, F. High-Resolution X-Ray Emission and X-Ray Absorption Spectroscopy. *Chem. Rev.* **2001**, *101* (6), 1779–1808. https://doi.org/10.1021/cr9900681.

(15) Groot, F. de. Multiplet Effects in X-Ray Spectroscopy. *Coordination Chemistry Reviews* **2005**, *249* (1–2), 31–63. https://doi.org/10.1016/j.ccr.2004.03.018.

(16) Gao, S.-P.; Pickard, C. J.; Payne, M. C.; Zhu, J.; Yuan, J. Theory of Core-Hole Effects in $1s$ Core-Level Spectroscopy of the First-Row Elements. *Phys. Rev. B* **2008**, *77* (11), 115122. https://doi.org/10.1103/PhysRevB.77.115122.

(17) Husek, J.; Cirri, A.; Biswas, S.; Baker, L. R. Surface Electron Dynamics in Hematite (α-$Fe_2O_3$): Correlation between Ultrafast Surface Electron Trapping and Small Polaron Formation. *Chem. Sci.* **2017**, *8* (12), 8170–8178. https://doi.org/10.1039/C7SC02826A.

(18) Vura-Weis, J.; Jiang, C.-M.; Liu, C.; Gao, H.; Lucas, J. M.; de Groot, F. M. F.; Yang, P.; Alivisatos, A. P.; Leone, S. R. Femtosecond $M_{2,3}$ -Edge Spectroscopy of Transition-Metal Oxides: Photoinduced Oxidation State Change in α-$Fe_2O_3$. *J. Phys. Chem. Lett.* **2013**, *4* (21), 3667–3671. https://doi.org/10.1021/jz401997d.

(19) Biswas, S.; Wallentine, S.; Bandaranayake, S.; Baker, L. R. Controlling Polaron Formation at Hematite Surfaces by Molecular Functionalization Probed by XUV Reflection-Absorption Spectroscopy. *J. Chem. Phys.* **2019**, *151* (10), 104701. https://doi.org/10.1063/1.5115163.

(20) Porter, I. J.; Cushing, S. K.; Carneiro, L. M.; Lee, A.; Ondry, J. C.; Dahl, J. C.; Chang, H.-T.; Alivisatos, A. P.; Leone, S. R. Photoexcited Small Polaron Formation in Goethite (α-FeOOH) Nanorods Probed by Transient Extreme Ultraviolet Spectroscopy. *J. Phys. Chem. Lett.* **2018**, *9* (14), 4120–4124. https://doi.org/10.1021/acs.jpclett.8b01525.

(21) Gedamu Tamirat, A.; Rick, J.; Aregahegn Dubale, A.; Su, W.-N.; Hwang, B.-J. Using Hematite for Photoelectrochemical Water Splitting: A Review of Current Progress and Challenges. *Nanoscale Horizons* **2016**, *1* (4), 243–267. https://doi.org/10.1039/C5NH00098J.

(22) Zandi, O.; Hamann, T. W. The Potential versus Current State of Water Splitting with Hematite. *Phys. Chem. Chem. Phys.* **2015**, *17* (35), 22485–22503. https://doi.org/10.1039/C5CP04267D.

(23) Sivula, K.; Le Formal, F.; Grätzel, M. Solar Water Splitting: Progress Using Hematite (α-Fe2O3) Photoelectrodes. *ChemSusChem* **2011**, *4* (4), 432–449. https://doi.org/10.1002/cssc.201000416.

(24) Bandaranayake, S.; Hruska, E.; Londo, S.; Biswas, S.; Baker, L. R. Small Polarons and Surface Defects in Metal Oxide Photocatalysts Studied Using XUV Reflection–Absorption Spectroscopy. *J. Phys. Chem. C* **2020**, acs.jpcc.0c07047. https://doi.org/10.1021/acs.jpcc.0c07047.

(25) Biswas, S.; Husek, J.; Londo, S.; Baker, L. R. Highly Localized Charge Transfer Excitons in Metal Oxide Semiconductors. *Nano Lett.* **2018**, *18* (2), 1228–1233. https://doi.org/10.1021/acs.nanolett.7b04818.

(26) Stoneham, A. M.; Gavartin, J.; Shluger, A. L.; Kimmel, A. V.; Ramo, D. M.; Rønnow, H. M.; Aeppli, G.; Renner, C. Trapping, Self-Trapping and the Polaron Family. *J. Phys.: Condens. Matter* **2007**, *19* (25), 255208. https://doi.org/10.1088/0953-8984/19/25/255208.

(27) Rettie, A. J. E.; Chemelewski, W. D.; Emin, D.; Mullins, C. B. Unravelling Small-Polaron Transport in Metal Oxide Photoelectrodes. *J. Phys. Chem. Lett.* **2016**, *7* (3), 471–479. https://doi.org/10.1021/acs.jpclett.5b02143.

(28) Adelstein, N.; Neaton, J. B.; Asta, M.; De Jonghe, L. C. Density Functional Theory Based Calculation of Small-Polaron Mobility in Hematite. *Phys. Rev. B* **2014**, *89* (24), 245115. https://doi.org/10.1103/PhysRevB.89.245115.

(29) Biswas, S.; Husek, J.; Londo, S.; Baker, L. R. Ultrafast Electron Trapping and Defect-Mediated Recombination in NiO Probed by Femtosecond Extreme Ultraviolet Reflection–Absorption Spectroscopy. *J. Phys. Chem. Lett.* **2018**, *9* (17), 5047–5054. https://doi.org/10.1021/acs.jpclett.8b01865.

(30) Stavitski, E.; de Groot, F. M. F. The CTM4XAS Program for EELS and XAS Spectral Shape Analysis of Transition Metal L Edges. *Micron* **2010**, *41* (7), 687–694. https://doi.org/10.1016/j.micron.2010.06.005.

(31) Cannelli, O.; Colonna, N.; Puppin, M.; Rossi, T. C.; Kinschel, D.; Leroy, L. M. D.; Löffler, J.; Budarz, J. M.; March, A. M.; Doumy, G.; Al Haddad, A.; Tu, M.-F.; Kumagai, Y.; Walko, D.; Smolentsev, G.; Krieg, F.; Boehme, S. C.; Kovalenko, M. V.; Chergui, M.; Mancini, G. F. Quantifying Photoinduced Polaronic Distortions in Inorganic Lead Halide Perovskite Nanocrystals. *J. Am. Chem. Soc.* **2021**, *143* (24), 9048–9059. https://doi.org/10.1021/jacs.1c02403.

(32) Soldatov, M. A.; Martini, A.; Bugaev, A. L.; Pankin, I.; Medvedev, P. V.; Guda, A. A.; Aboraia, A. M.; Podkovyrina, Y. S.; Budnyk, A. P.; Soldatov, A. A.; Lamberti, C. The Insights from X-Ray Absorption Spectroscopy into the Local Atomic Structure and Chemical Bonding of Metal–Organic Frameworks. *Polyhedron* **2018**, *155*, 232–253. https://doi.org/10.1016/j.poly.2018.08.004.

(33) Guda, S. A.; Guda, A. A.; Soldatov, M. A.; Lomachenko, K. A.; Bugaev, A. L.; Lamberti, C.; Gawelda, W.; Bressler, C.; Smolentsev, G.; Soldatov, A. V.; Joly, Y. Optimized Finite Difference Method for the Full-Potential XANES Simulations: Application to Molecular Adsorption Geometries in MOFs and Metal–Ligand Intersystem Crossing Transients. *J. Chem. Theory Comput.* **2015**, *11* (9), 4512–4521. https://doi.org/10.1021/acs.jctc.5b00327.

(34) Vinson, J.; Jach, T.; Müller, M.; Unterumsberger, R.; Beckhoff, B. Quasiparticle Lifetime Broadening in Resonant X-Ray Scattering of ${\mathrm{NH}}_{4}{\mathrm{NO}}_{3}$. *Phys. Rev. B* **2016**, *94* (3), 035163. https://doi.org/10.1103/PhysRevB.94.035163.

(35) Li, L.; Zhang, R.; Vinson, J.; Shirley, E. L.; Greeley, J. P.; Guest, J. R.; Chan, M. K. Y. Imaging Catalytic Activation of CO2 on Cu2O (110): A First-Principles Study. *Chem. Mater.* **2018**, *30* (6), 1912–1923. https://doi.org/10.1021/acs.chemmater.7b04803.

(36) Woicik, J. C.; Weiland, C.; Rumaiz, A. K.; Brumbach, M.; Quackenbush, N. F.; Ablett, J. M.; Shirley, E. L. Revealing Excitonic Processes and Chemical Bonding in $\mathrm{Mo}{\mathrm{S}}_{2}$ by X-Ray Spectroscopy. *Phys. Rev. B* **2018**, *98* (11), 115149. https://doi.org/10.1103/PhysRevB.98.115149.

(37) Geondzhian, A.; Gilmore, K. Demonstration of Resonant Inelastic X-Ray Scattering as a Probe of Exciton-Phonon Coupling. *Phys. Rev. B* **2018**, *98* (21), 214305. https://doi.org/10.1103/PhysRevB.98.214305.

(38) Attar, A. R.; Chang, H.-T.; Britz, A.; Zhang, X.; Lin, M.-F.; Krishnamoorthy, A.; Linker, T.; Fritz, D.; Neumark, D. M.; Kalia, R. K.; Nakano, A.; Ajayan, P.; Vashishta, P.; Bergmann, U.; Leone, S. R. Simultaneous Observation of Carrier-Specific Redistribution and Coherent Lattice Dynamics in 2H-MoTe2 with Femtosecond Core-Level Spectroscopy. *ACS Nano* **2020**, *14* (11), 15829–15840. https://doi.org/10.1021/acsnano.0c06988.

(39) Vinson, J.; Rehr, J. J.; Kas, J. J.; Shirley, E. L. Bethe-Salpeter Equation Calculations of Core Excitation Spectra. *Phys. Rev. B* **2011**, *83* (11), 115106. https://doi.org/10.1103/PhysRevB.83.115106.

(40) Gilmore, K.; Vinson, J.; Shirley, E. L.; Prendergast, D.; Pemmaraju, C. D.; Kas, J. J.; Vila, F. D.; Rehr, J. J. Efficient Implementation of Core-Excitation Bethe–Salpeter Equation Calculations. *Computer Physics Communications* **2015**, *197*, 109–117. https://doi.org/10.1016/j.cpc.2015.08.014.

(41) Kvashnina, K. O.; Butorin, S. M. High-Energy Resolution X-Ray Spectroscopy at Actinide M4,5 and Ligand K Edges: What We Know, What We Want to Know, and What We Can Know. *Chem. Commun.* **2022**, *58* (3), 327–342. https://doi.org/10.1039/D1CC04851A.

(42) Nattino, F.; Marzari, N. Operando XANES from First-Principles and Its Application to Iridium Oxide. *Phys. Chem. Chem. Phys.* **2020**, *22* (19), 10807–10818. https://doi.org/10.1039/C9CP06726D.

(43) OceanDoc.Pdf.

(44) Pham, T. D.; Deskins, N. A. Efficient Method for Modeling Polarons Using Electronic Structure Methods. *J. Chem. Theory Comput.* **2020**, *16* (8), 5264–5278. https://doi.org/10.1021/acs.jctc.0c00374.

(45) Shibuya, T.; Yasuoka, K.; Mirbt, S.; Sanyal, B. A Systematic Study of Polarons Due to Oxygen Vacancy Formation at the Rutile $TiO_2$ (110) Surface by GGA + *U* and HSE06 Methods. *J. Phys.: Condens. Matter* **2012**, *24* (43), 435504. https://doi.org/10.1088/0953-8984/24/43/435504.

(46) Deskins, N. A.; Dupuis, M. Electron Transport via Polaron Hopping in Bulk Ti O 2 : A Density Functional Theory Characterization. *Phys. Rev. B* **2007**, *75* (19), 195212. https://doi.org/10.1103/PhysRevB.75.195212.

(47) Shelton, J. L.; Knowles, K. E. Thermally Activated Optical Absorption into Polaronic States in Hematite. *J. Phys. Chem. Lett.* **2021**, *12* (13), 3343–3351. https://doi.org/10.1021/acs.jpclett.0c03751.

(48) Smyth, J. R.; Jacobsen, S. D.; Hazen, R. M. Comparative Crystal Chemistry of Dense Oxide Minerals. *Reviews in Mineralogy and Geochemistry* **2000**, *41* (1), 157–186. https://doi.org/10.2138/rmg.2000.41.6.

(49) Liu, H.; Michelsen, J. M.; Klein, I. M.; Cushing, S. K. Measuring Photoexcited Electron and Hole Dynamics in ZnTe and Modeling Excited State Core-Valence Effects in Transient XUV Reflection Spectroscopy. *arXiv:2108.02262 [cond-mat, physics:physics]* **2021**.

**_Ab Initio_ Prediction of Excited State and Polaron Effects in Transient XUV Measurements of α-$Fe_2O_3$**

Isabel M. Klein,[1] Hanzhe Liu,[1] Danika Nimlos,[1] Alex Krotz,[1] Scott K. Cushing[1,†]

[1]Division of Chemistry and Chemical Engineering, California Institute of Technology, Pasadena, CA 91125, USA.

[†]Corresponding author. Email: scushing@caltech.edu

**Contents:**

## 1. Computational Methods

Geometry optimization and DFT calculations were performed with the Quantum ESPRESSO package using Perdew-Burke-Ernzerhof (PBE) pseudopotentials under the generalized gradient approximation (GGA). The associated ground state wavefunctions were constructed using a plane wave basis set with components up to a kinetic energy cutoff of 250 Ry. Reciprocal space was sampled using an 4x4x4 Gamma-centered mesh with a 0.02 eV Gaussian smearing of orbital occupancies. DFT simulations were performed on a unit cell of $\alpha$-$Fe_2O_3$ containing 10 atoms, measuring 10.24 Å along the *a*-, *b*-and *c*- directions. Self-consistent calculations were performed to a convergence of $10^{-6}$ eV/atom and forces on ions under $10^{-3}$ eV/Å. DFT-BSE calculations were conducted using the same parameters. Additionally, an 8x8x8 k-point grid for the screening mesh, 100 bands, a dielectric constant of 25, a cutoff radius of 4.5 Bohr, and a band gap scissor correction of 3 eV were used in the DFT-BSE calculations of the XUV absorption spectra.[1,2]

## 2. Structural Data for Calculations

a. Calculations for both the ground state and the state-filling excited state of α-$Fe_2O_3$ were performed on the following unit cell.

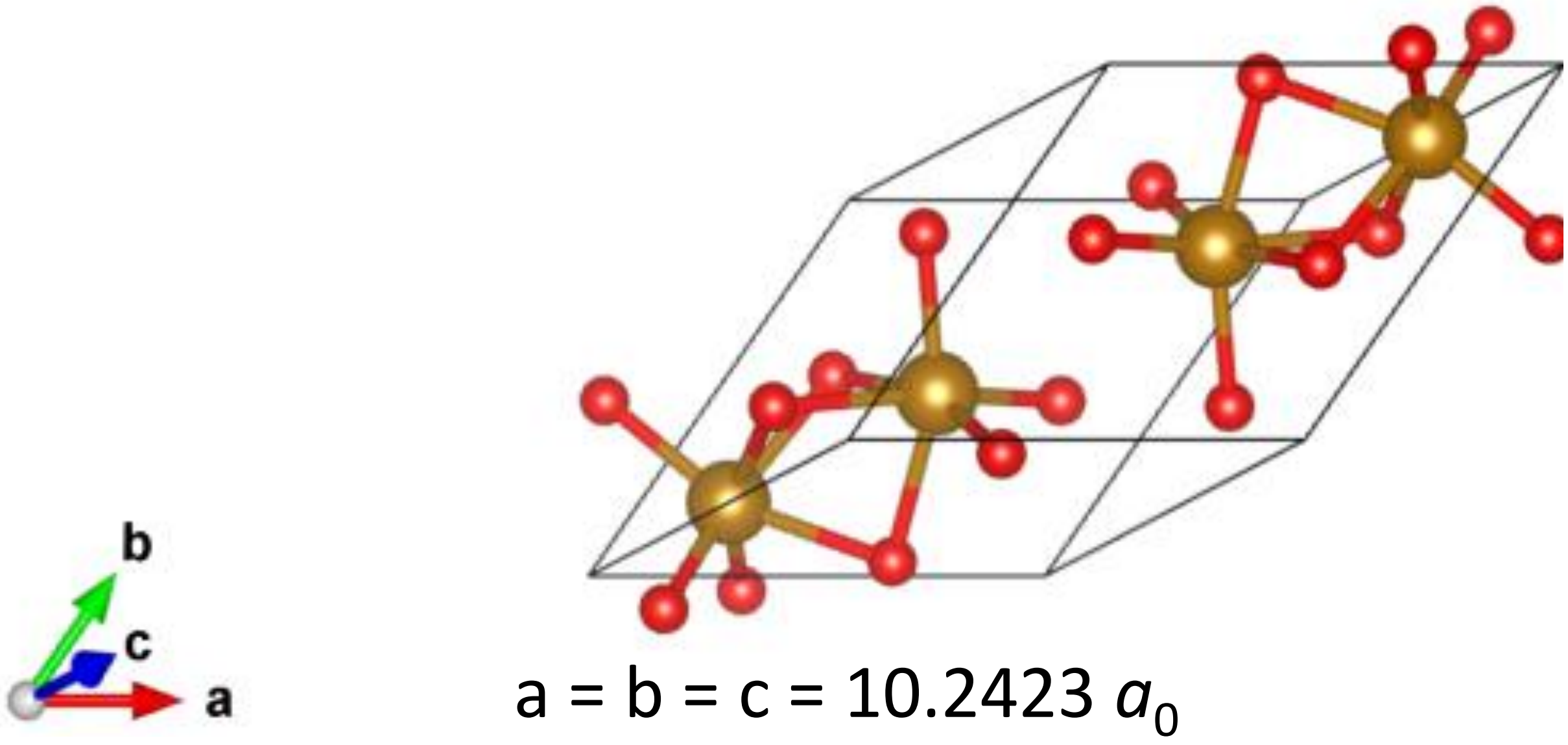

Figure S1. Unit Cell for Ground State and State Filling Excited State OCEAN Calculations.

Primitive Vectors

| | | |
|---|---|---|
| 1.0000000000 | 0.0000000000 | 0.0000000000 |
| 0.5695664700 | 0.8219452757 | 0.0000000000 |
| 0.5695664700 | 0.2982686482 | 0.7659176521 |

Reduced coordinates, ( x, y, z ), of the all n-atoms in the cell were

| | | | |
|---|---|---|---|
| Fe1 | 0.1433915 | 0.1433904 | 0.1433906 |
| Fe1 | 0.8566085 | 0.8566096 | 0.8566094 |
| Fe2 | 0.3566087 | 0.3566104 | 0.3566106 |
| Fe2 | 0.6433913 | 0.6433896 | 0.6433894 |
| O | 0.7500005 | 0.4472258 | 0.0527735 |
| O | 0.9472262 | 0.2499995 | 0.5527741 |
| O | 0.4472260 | 0.0527734 | 0.7500002 |
| O | 0.2499995 | 0.5527742 | 0.9472265 |
| O | 0.0527738 | 0.7500005 | 0.4472259 |
| O | 0.5527740 | 0.9472266 | 0.2499998 |

b. Calculations for the polaron excited state of $\alpha$-$Fe_2O_3$ were performed using the following unit cell, arrived at by the procedure described in the main text. The Fe2 site is used for the polaron state X-ray site and is highlighted in yellow.

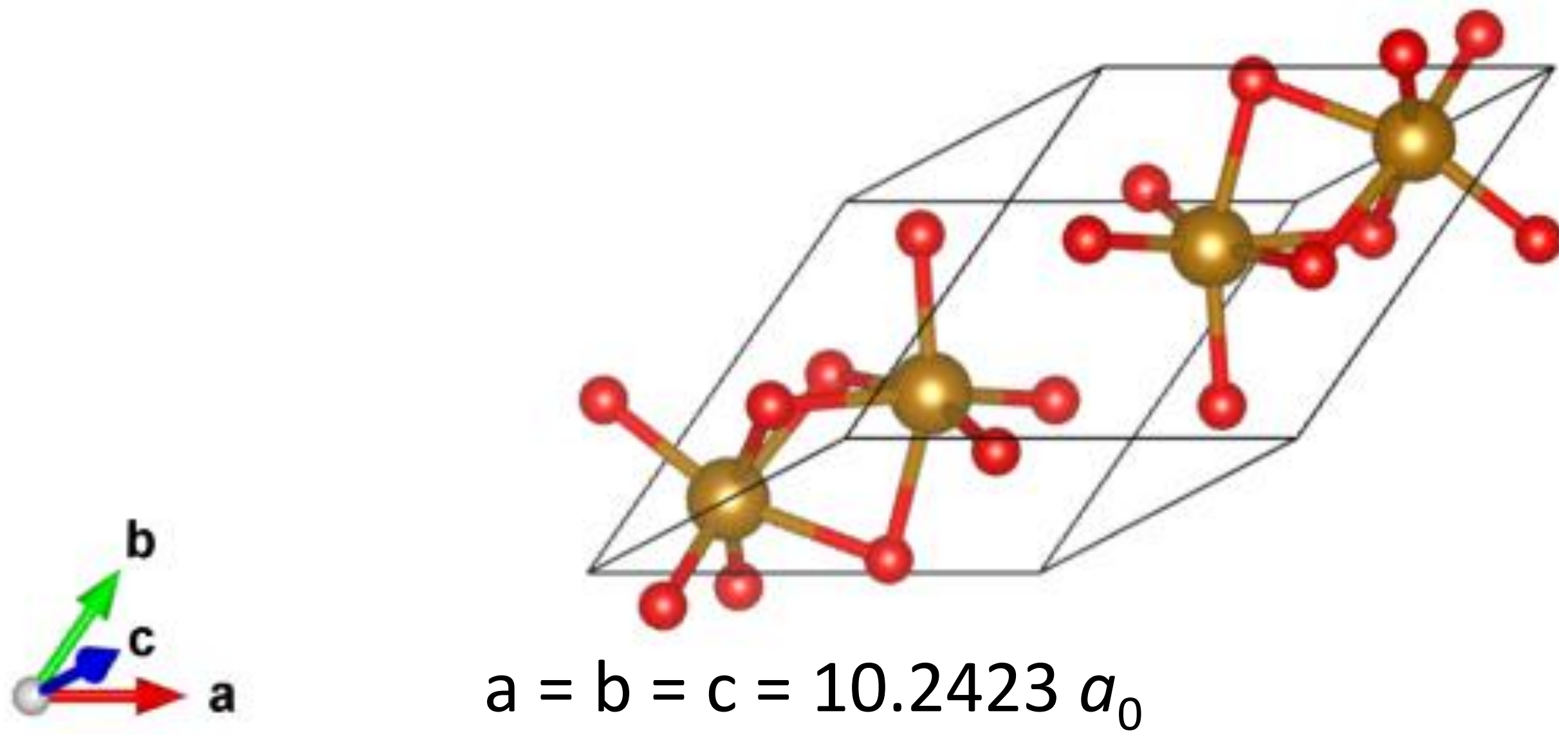

Figure S2. Unit Cell for Polaron Excited State OCEAN Calculations.

Primitive Vectors

| | | |
|---|---|---|
| 1.0000000000 | 0.0000000000 | 0.0000000000 |
| 0.5695664700 | 0.8219452757 | 0.0000000000 |
| 0.5695664700 | 0.2982686482 | 0.7659176521 |

Reduced Coordinates, ( x, y, z ), of the all natom atoms in the cell

| | | | |
|---|---|---|---|
| Fe1 | 0.1452347 | 0.1454904 | 0.1451297 |
| Fe1 | 0.8549303 | 0.8545129 | 0.8546109 |
| Fe2 | 0.3544541 | 0.3549705 | 0.3547903 |
| Fe2 | 0.6450367 | 0.6458038 | 0.6448248 |
| O | 0.7498179 | 0.4453231 | 0.0551344 |
| O | 0.9451309 | 0.2494991 | 0.5553065 |
| O | 0.4449728 | 0.0547092 | 0.7500116 |
| O | 0.2500428 | 0.5549325 | 0.9452755 |
| O | 0.0553823 | 0.7494405 | 0.4452824 |
| O | 0.5549975 | 0.9453182 | 0.2496338 |

c. Calculations for the thermally expanded excited state of α-$Fe_2O_3$ were performed on the unit cell below. The specific change to the scaling of the primitive vectors that models the thermal expansion is highlighted in yellow.

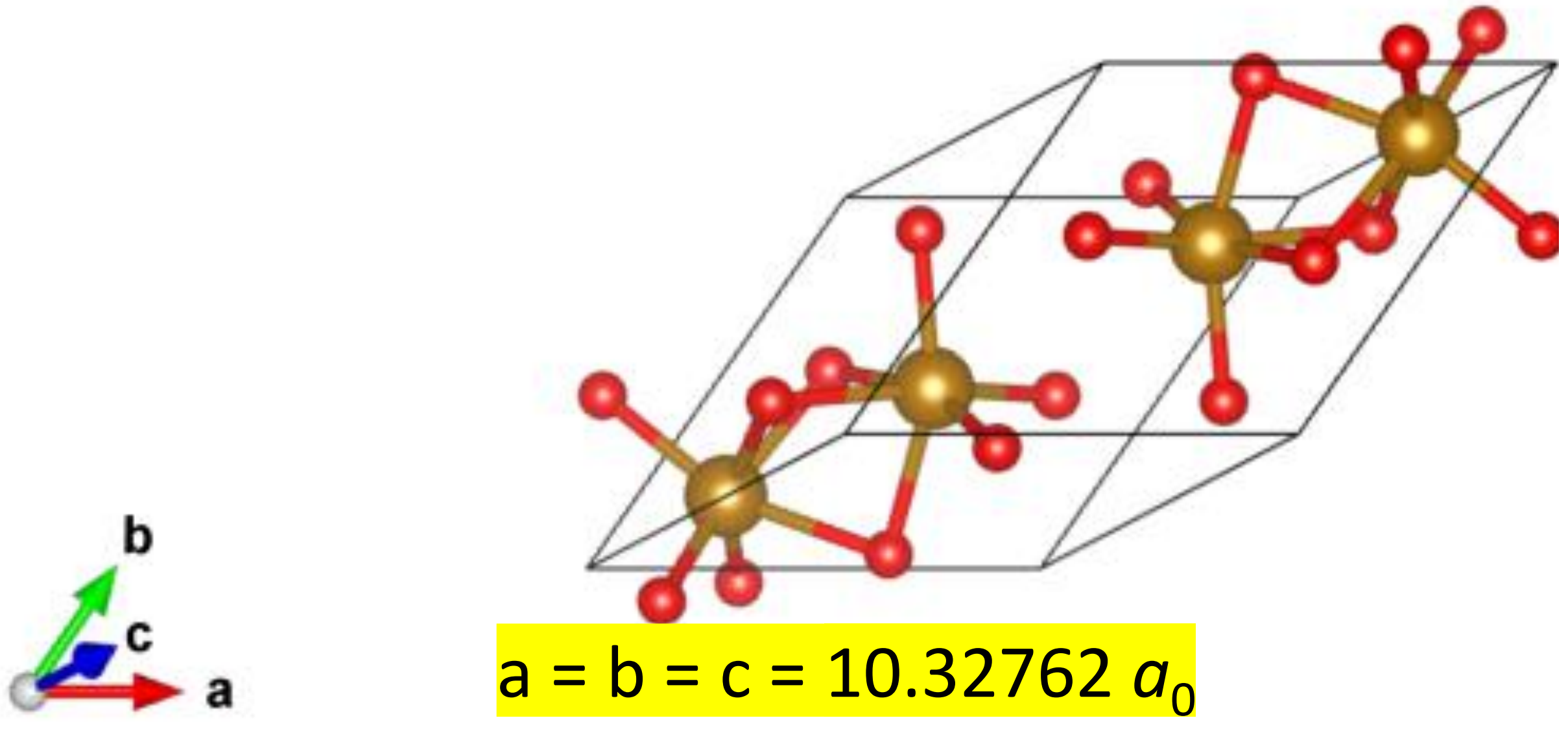

$a = b = c = 10.32762\ a_0$

Figure S3. Unit Cell for Thermally Expanded Excited State OCEAN Calculations.

Primitive Vectors

| | | |
|---|---|---|
| 1.0000000000 | 0.0000000000 | 0.0000000000 |
| 0.5695664700 | 0.8219452757 | 0.0000000000 |
| 0.5695664700 | 0.2982686482 | 0.7659176521 |

Reduced Coordinates, ( x, y, z ), of the all natom atoms in the cell

| | | | |
|---|---|---|---|
| Fe1 | 0.1433915 | 0.1433904 | 0.1433906 |
| Fe1 | 0.8566085 | 0.8566096 | 0.8566094 |
| Fe2 | 0.3566087 | 0.3566104 | 0.3566106 |
| Fe2 | 0.6433913 | 0.6433896 | 0.6433894 |
| O | 0.7500005 | 0.4472258 | 0.0527735 |
| O | 0.9472262 | 0.2499995 | 0.5527741 |
| O | 0.4472260 | 0.0527734 | 0.7500002 |
| O | 0.2499995 | 0.5527742 | 0.9472265 |
| O | 0.0527738 | 0.7500005 | 0.4472259 |
| O | 0.5527740 | 0.9472266 | 0.2499998 |

d. Comparisons between each excited state and the ground state for the Fe-Fe distance and Fe-O distance are given in Table S1, along with the percent change from the ground state.

Table S1. Bond Distances and Change from the Ground State for each Excite State.

| **State** | **Fe-Fe Distance (Å)** | **Percent Change from Ground (%)** | **Fe-O Distance (Å)** | **Percent Change from Ground (%)** |
|---|---|---|---|---|
| Ground | 2.92755 | n/a | 1.94192 | n/a |
| Charge Transfer | 2.92755 | 0 | 1.94192 | 0 |
| Polaron | 2.87585 | - 1.77 | 1.95016 | + 0.4 |
| Thermal (350 K) | 2.95194 | + 0.83 | 1.95810 | + 0.83 |

## 3. Ground State Calculations

a. Initial DFT calculations on the ground state of α-$Fe_2O_3$ were performed to verify computational parameters used in this study. The OCEAN computational program is not yet able to use hybrid pseudopotentials, and as such it was important to compare the band structure and DOS shown here to previous computational work.[2] Partial DOS with orbital breakdowns are shown to fully understand the DOS and the excitations calculated in the OCEAN calculations. As commented earlier, a scissor shift was used for to correct the band gap because hybrid pseudopotentials could not be used. Since only the conduction band states are needed for the final calculations, the calculation can still be accurate even without a hybrid functional. Accordingly, adjusting the U parameter in the DFT+U calculations had little effect on the final XUV absorption spectrum.

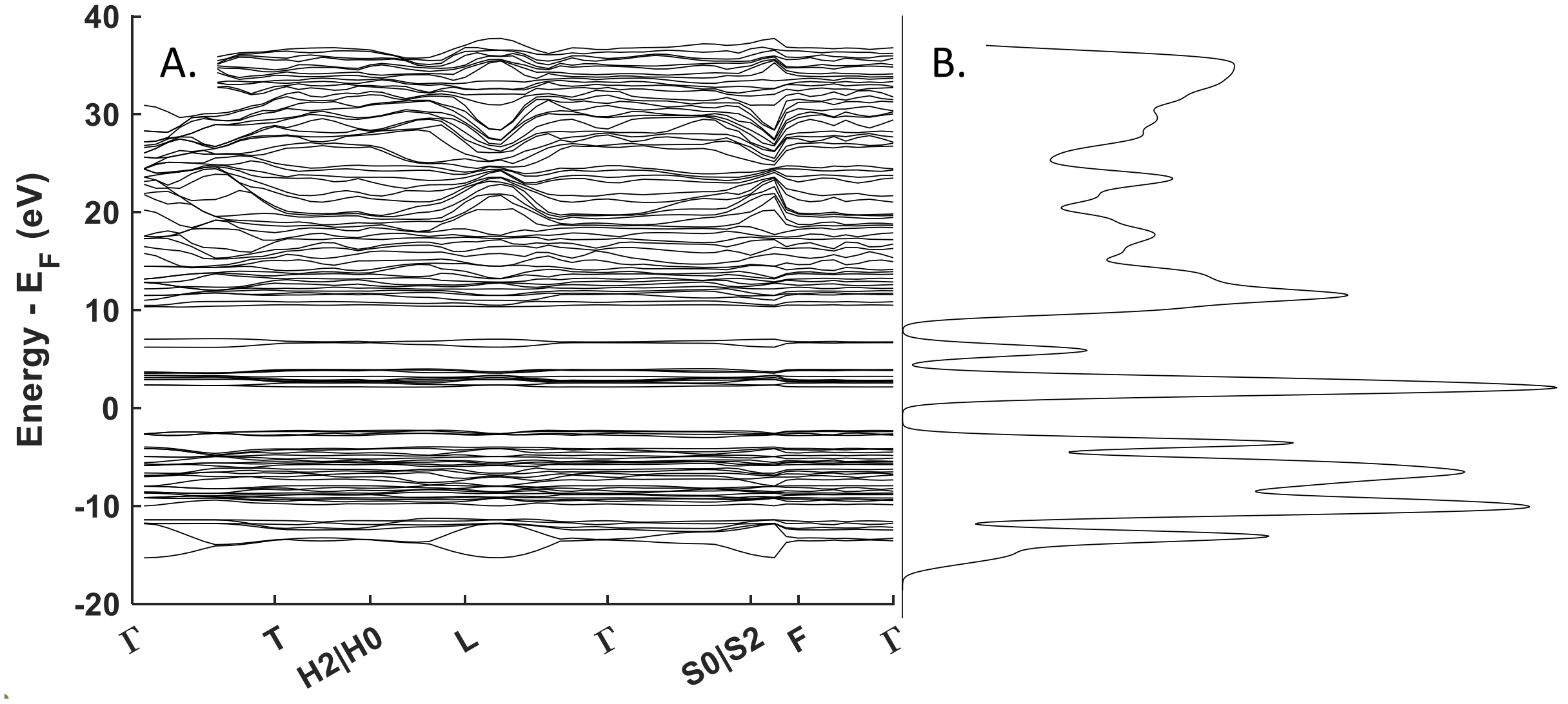

Figure S4. Ground State DFT Outputs for α-$Fe_2O_3$. (a) Full ground state band structure with the Fermi energy set to 0 eV. (b) Full ground state total density of states. No scissor shift has been included in these calculations.

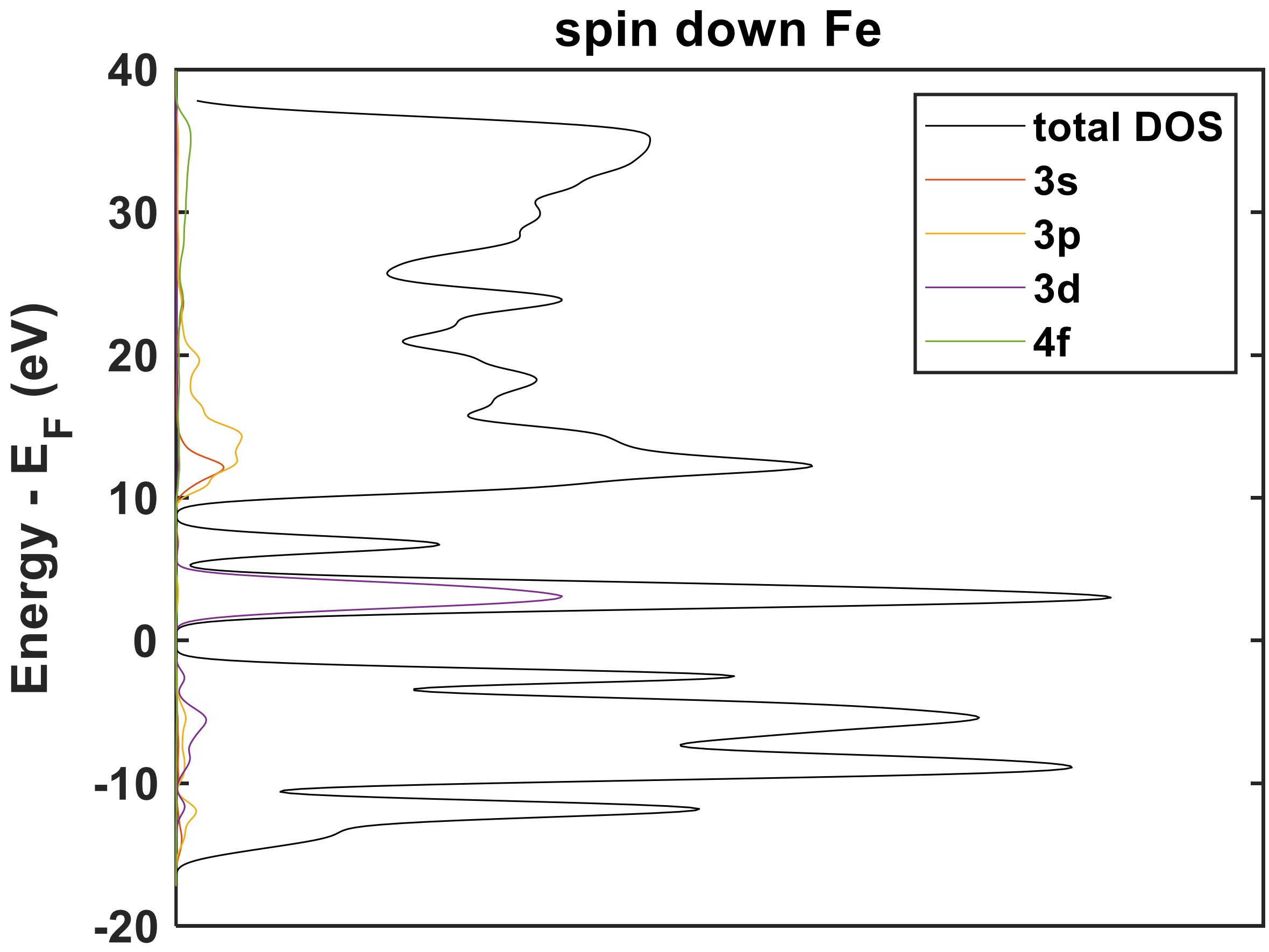

Figure S5. Partial Density of States for α-$Fe_2O_3$. Spin down Fe partial DOS decomposed into 3s, 3p, 3d, and 4f orbital contributions. No scissor shift has been included in these calculations.

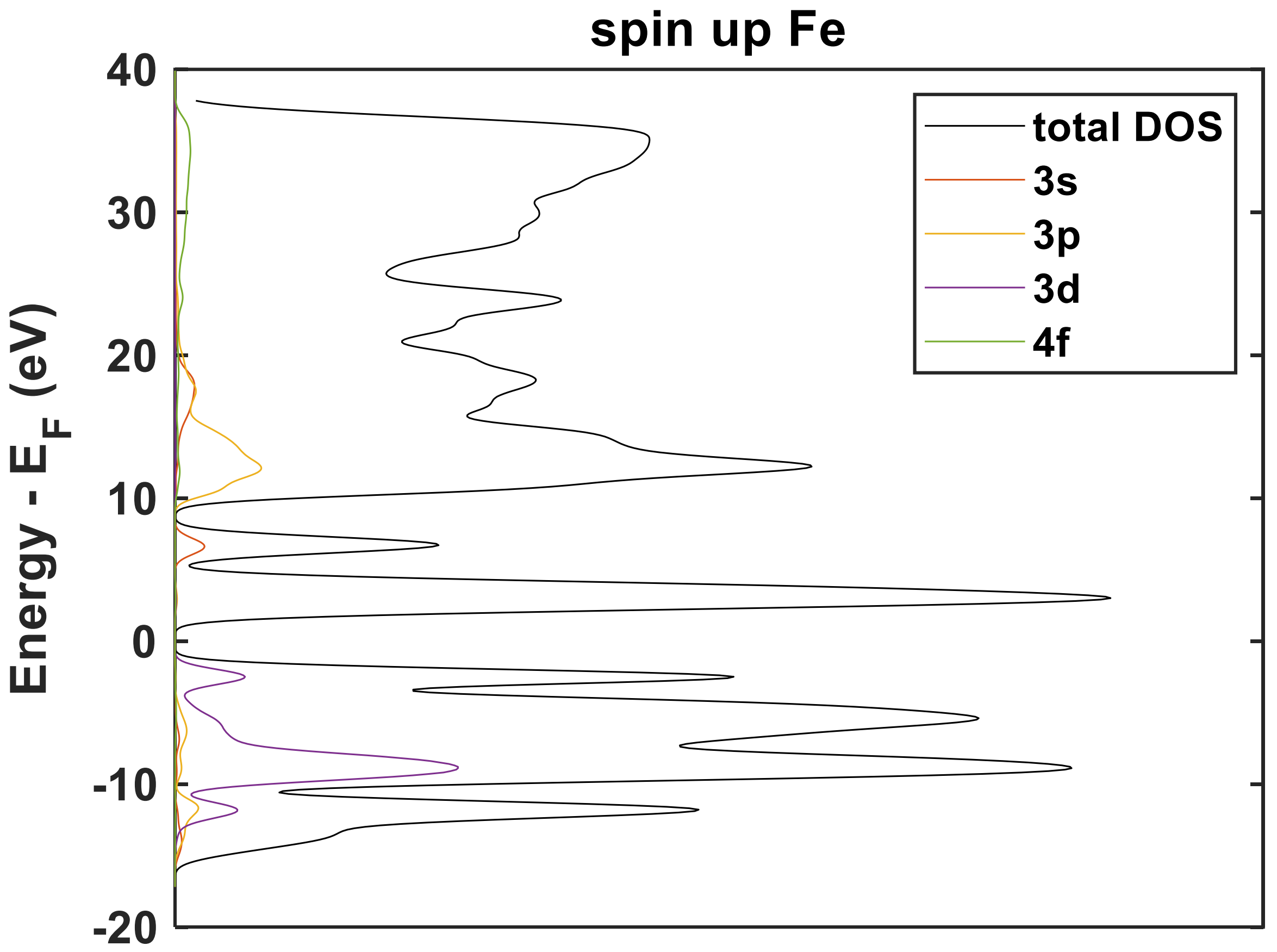

Figure S6. Partial Density of States for α-$Fe_2O_3$. Spin up Fe partial DOS decomposed into 3s, 3p, 3d, and 4f orbital contributions. No scissor shift has been included in these calculations.

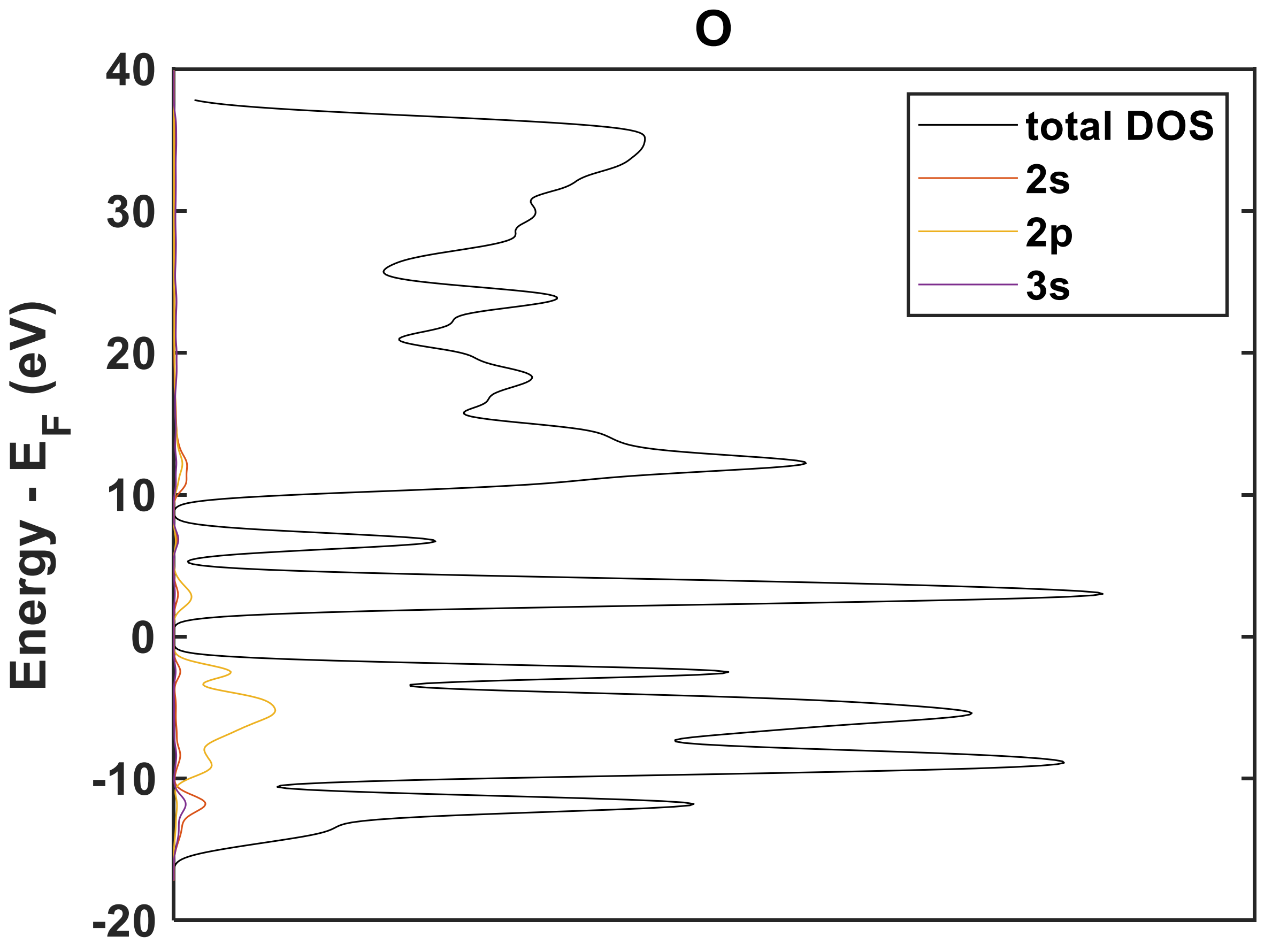

Figure S7. Partial Density of States for α-$Fe_2O_3$. O partial DOS decomposed into 2s, 2p, and 3p orbital contributions. No scissor shift has been included in these calculations.

b. The ground state XUV absorption was calculated and broadened by convoluting the calculated XUV absorption spectrum with an energy-dependent Gaussian, as described below (consistent with previous literature, as can be compared in Figure S8).[3,4] The calculated ground state spectrum was compared to the LMCT stick spectrum used in previous work.[3]

The matlab script for the broadening is as follows:

$$G(E,w,E_o)=\frac{e^{-(\frac{E-E_0}{\omega\sqrt{2}})^2}}{\omega\sqrt{2\pi}}$$

Fe2O3_calc = initial calculated absorption spectrum
Fe2O3_ground_broad= final broadened calculated absorption spectrum
Fe2O3_broad = broadened calculated absorption spectrum
E(i) = 1 – 200 eV, 0.2 eV steps

```
for i=1:100
   if E(i) <= 58
         Fe2O3_broad=25*Fe2O3_calc(i)*G(E,0.5,E(i));
   elseif E(i) > 58 & E(i) <= 62
         Fe2O3_broad=300*Fe2O3_calc(i)*G(E,(i/200)^0.01 + 0.5,E(i));
   else
         Fe2O3_broad=200*Fe2O3_calc(i)*G(E, (i/200)^-2,E(i));
   end
   Fe2O3_ground_broad = Fe2O3_broad;
end
```

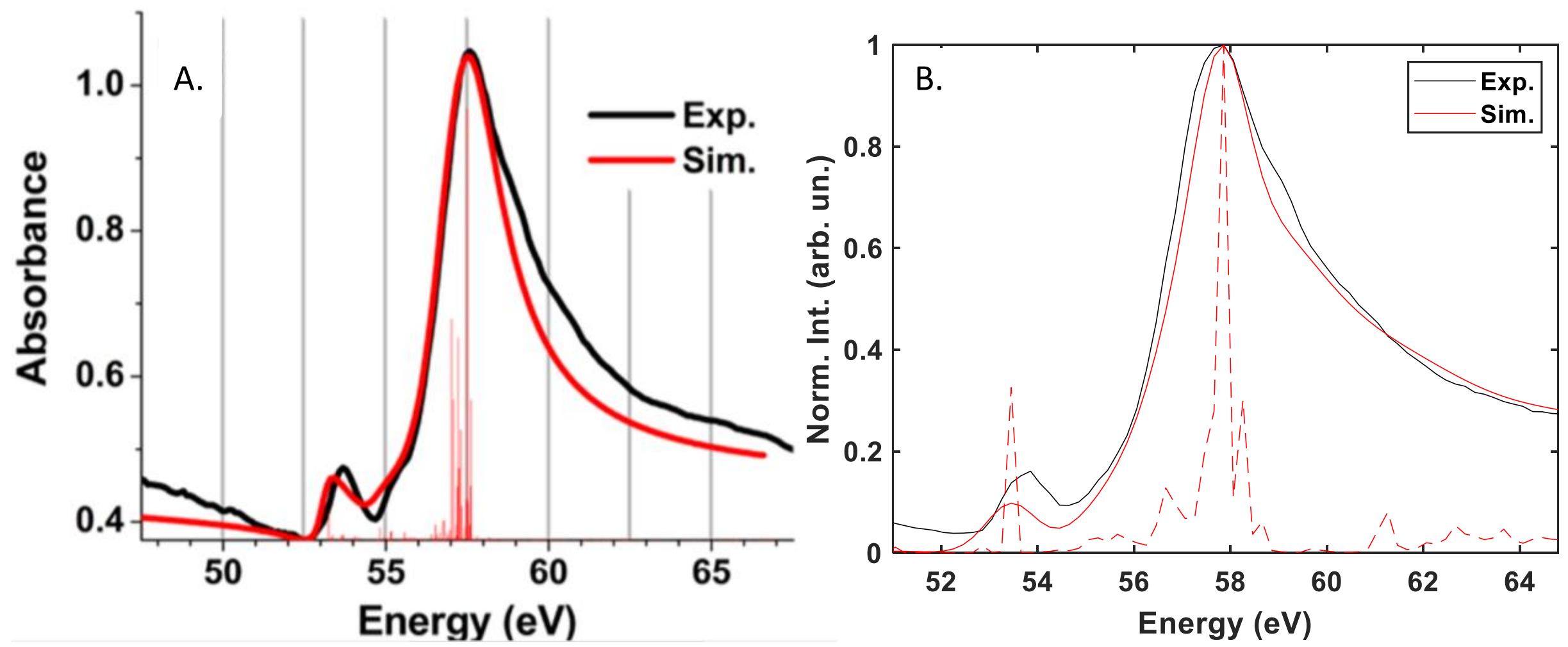

Figure S8. Comparison Between Previous Ligand Field Multiplet Simulations (a) and OCEAN Simulations (b) for XUV absorption spectrum of α-$Fe_2O_3$. Ligand field multiplet simulations are reproduced here from reference 3.

c. The ground state XUV core-valence exciton density was calculated for specific energy ranges and overlaid on the XUV absorption spectrum and band structure. To calculate the exciton density for these specific energy ranges, a GMRES calculation was used, restricting the energy to a 1 eV range. The range was changed to allow the entire region around the Fe $M_{2,3}$ edge under investigation to be calculated.

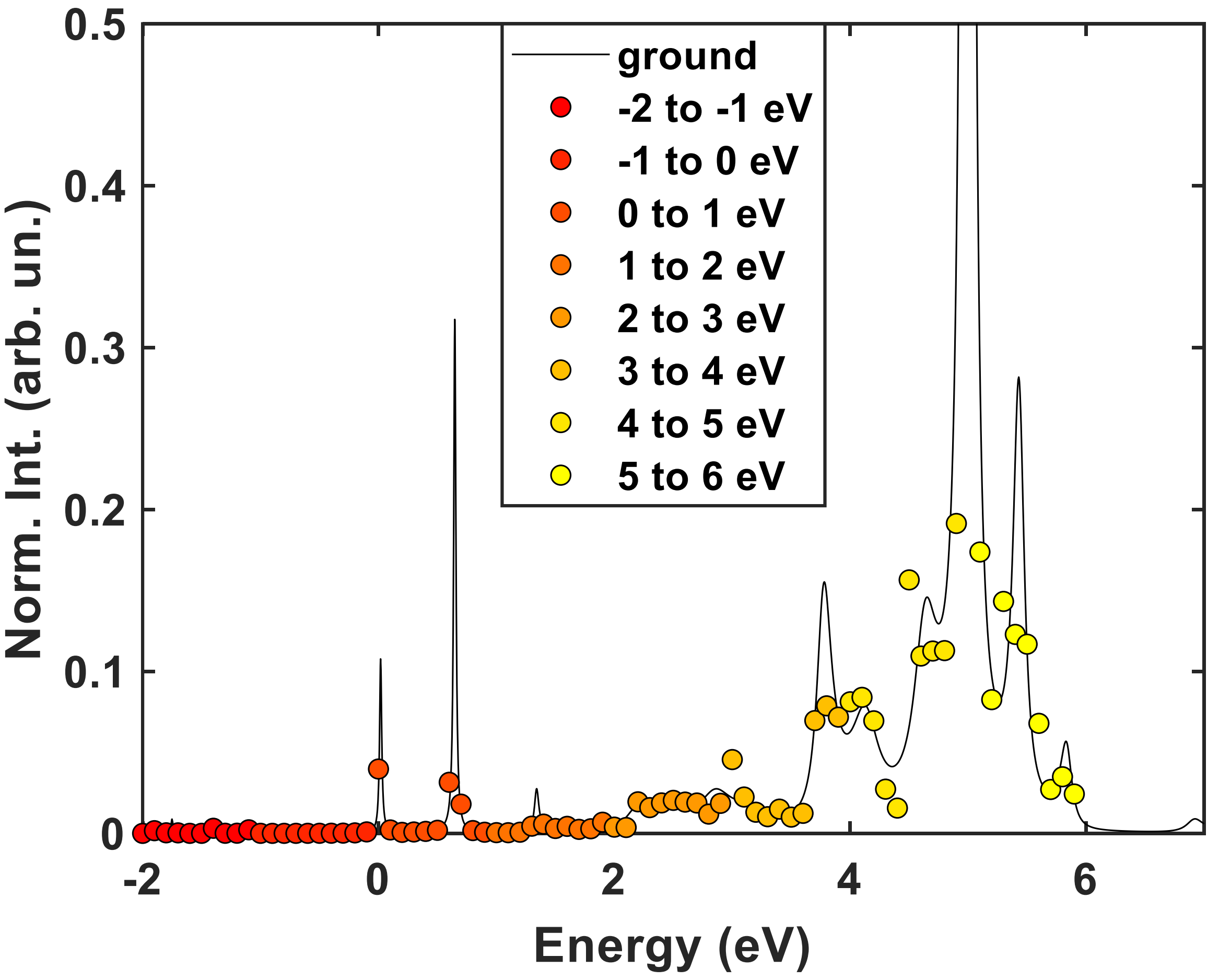

Figure S9. Energy-Resolved Exciton Densities Overlaid on the Calculated XUV Absorption Spectrum for α-$Fe_2O_3$. Exciton densities overlaid on the calculated XUV absorption spectrum around the Fe $M_{2,3}$ edge.

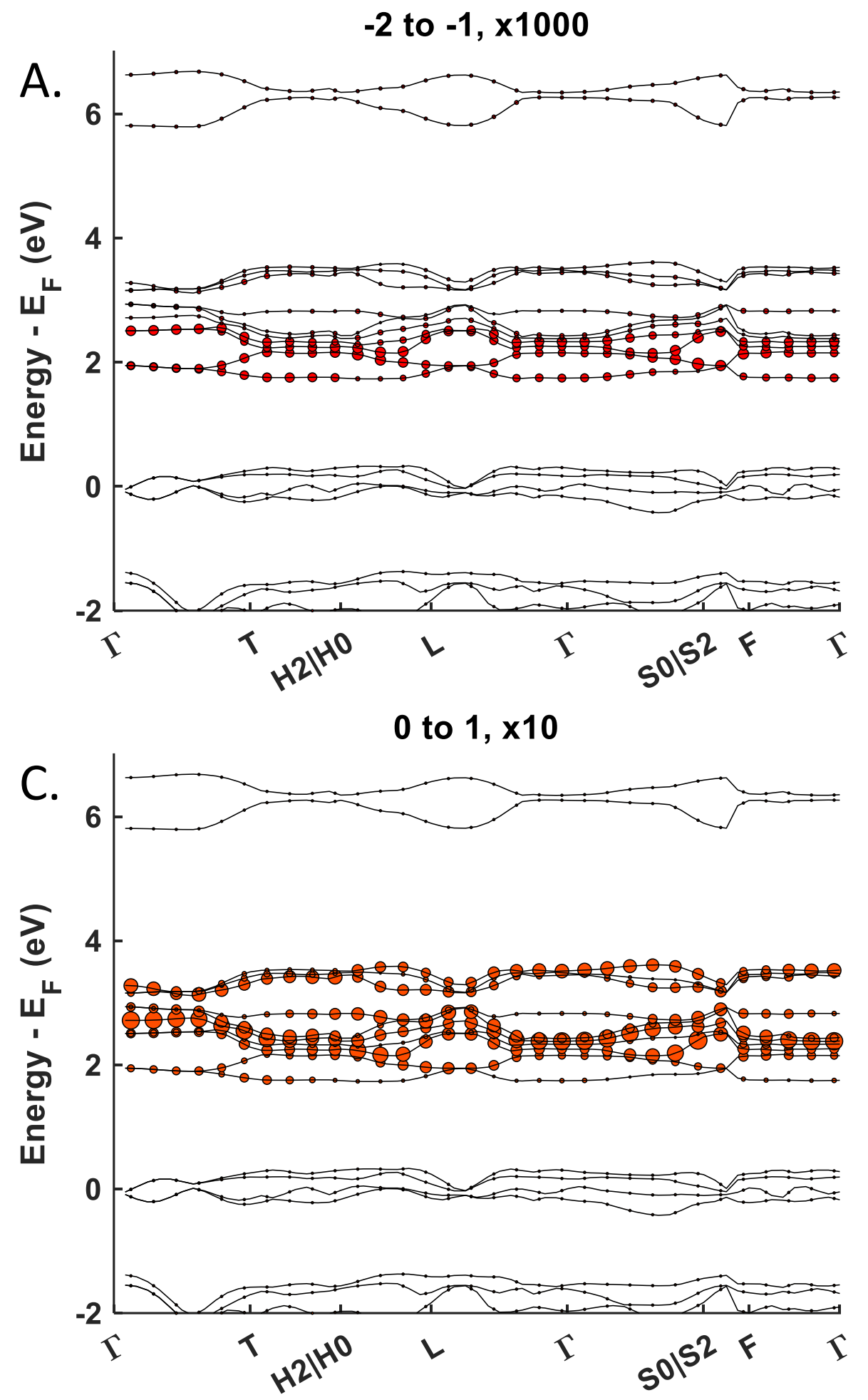

Figure S10. Energy-Resolved Exciton Densities Overlaid on the α-$Fe_2O_3$ Band Structure. Energy ranges between (a) -2 to -1 eV, (b) -1 to 0 eV, (c) 0 to 1 eV and (d) 1 to 2 around the Fe $M_{2,3}$ edge are shown. The necessary amplitude magnification is shown in the title of each band structure, demonstrating the relative strength of the exciton components in different energy ranges.

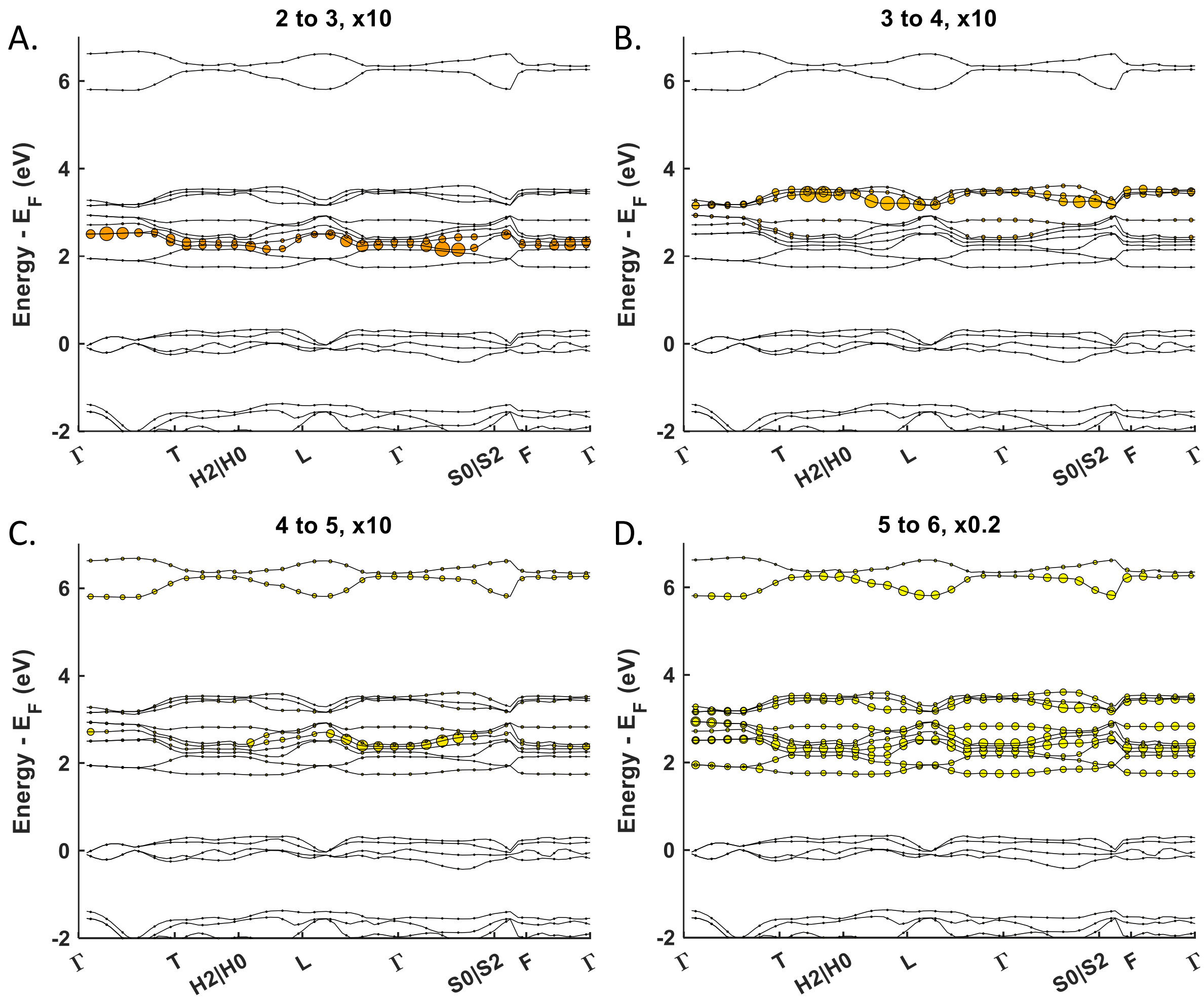

Figure S11. Energy-Resolved Exciton Densities Overlaid on the α-$Fe_2O_3$ Band Structure. Energy ranges between (a) 2 to 3 eV, (b) 3 to 4 eV, (c) 4 to 5 eV, and (d) 5 to 6 eV around the Fe $M_{2,3}$ edge are shown. The necessary amplitude magnification is shown in the title of each band structure, demonstrating the relative strength of the exciton components in different energy ranges.

## 4. Hamiltonian Decompositions

a. Comparisons between the total core-valence exciton density for the four states (ground, state-filling, polaron, and thermally expanded) modeled in this study.

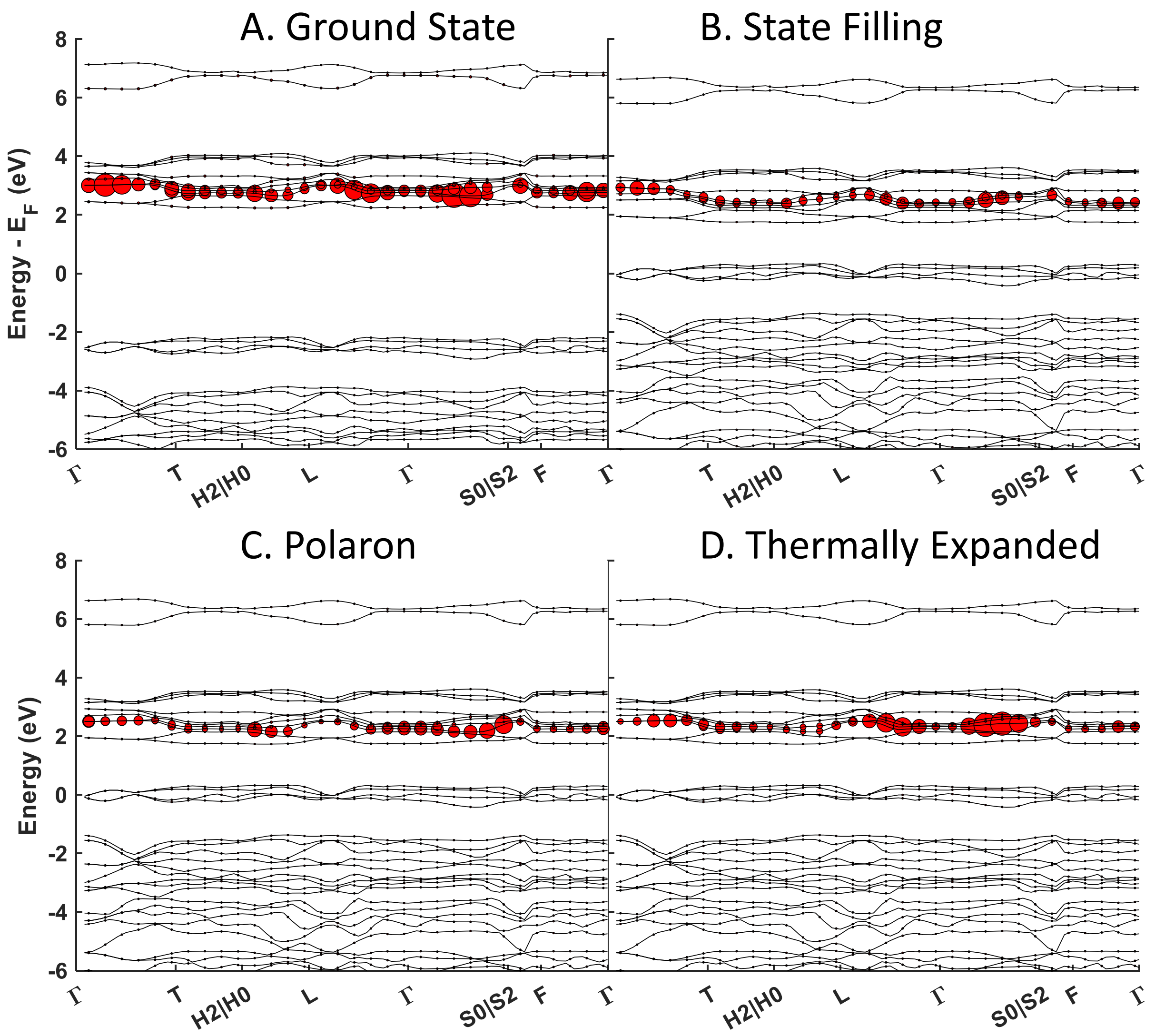

Figure S12. Total Core-Valence Exciton Densities. (a) Ground, (b) state filling, (c) polaron, and (d) thermally expanded states' total core-valence exciton densities overlaid on the band structure of α-$Fe_2O_3$.

b. Hamiltonian decomposition of the core-valence exciton for the ground state, state filling, polaron, and thermally expanded excited states. These decompositions are projected onto the band structure and are used to qualitatively compare trends between states, as well as quantify the differences between the excited states and the ground state Hamiltonian contributions. All contributions for one state are normalized relative to each other to facilitate direct comparison. The angular momentum, screening, and miltoplet effects are discussed in the main text. The high-order bubble and ladder many-body interactions relate to the formation and renormalization of the core-valence exciton.

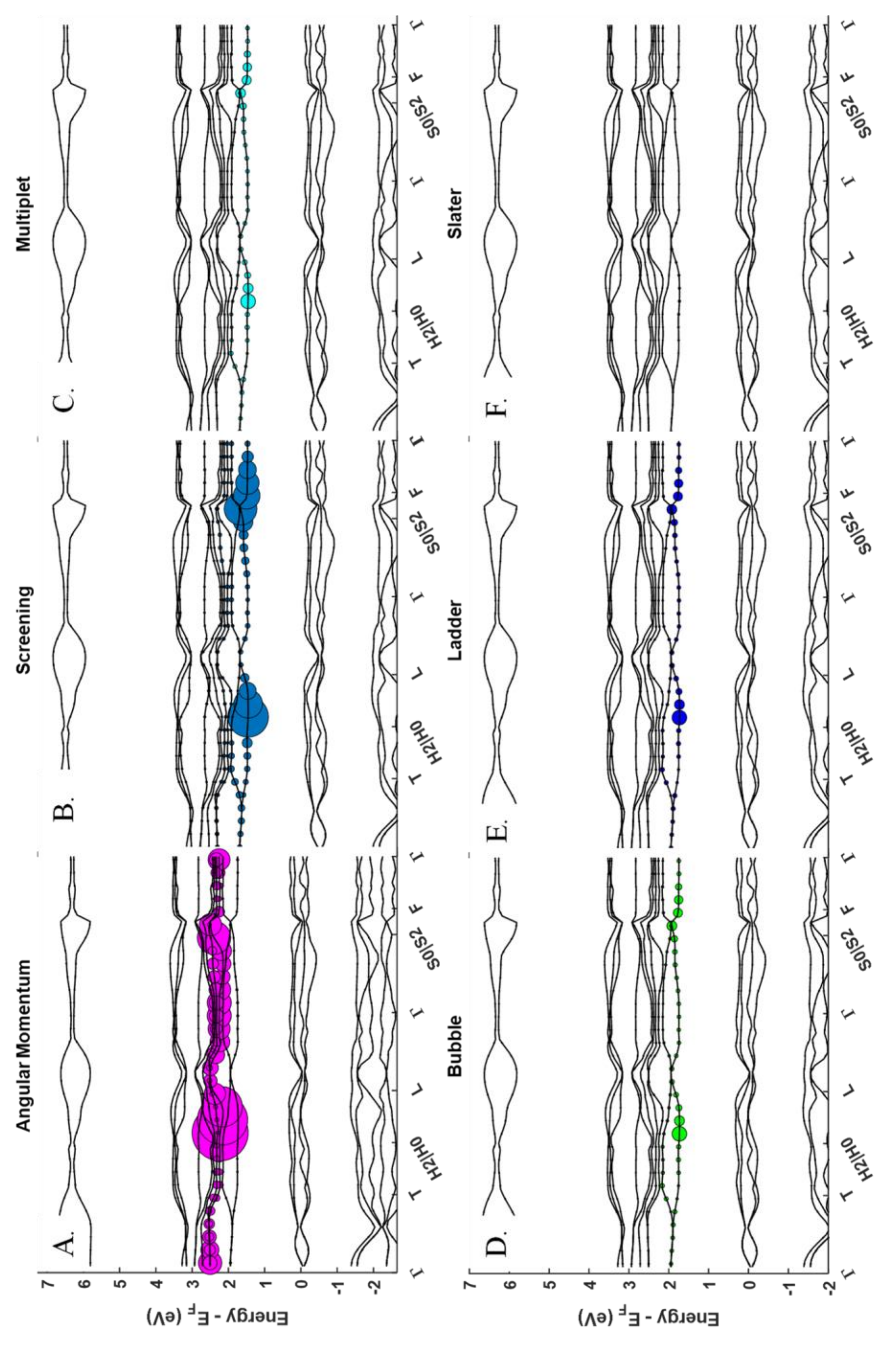

Figure S13. XUV transition Hamiltonian Decomposition of Core-Valence Exciton for the Ground State of α-$Fe_2O_3$. The (a) angular momentum, (b) core-hole screening, (c) atomic multiplet, (d) bubble, (e) ladder and (f) Slater core-valence exciton densities are calculated and normalized to determine the relative contribution of each of these processes to the overall XUV exciton and thus absorption spectrum. The angular momentum, multiplet and screening contributions are included in the main text.

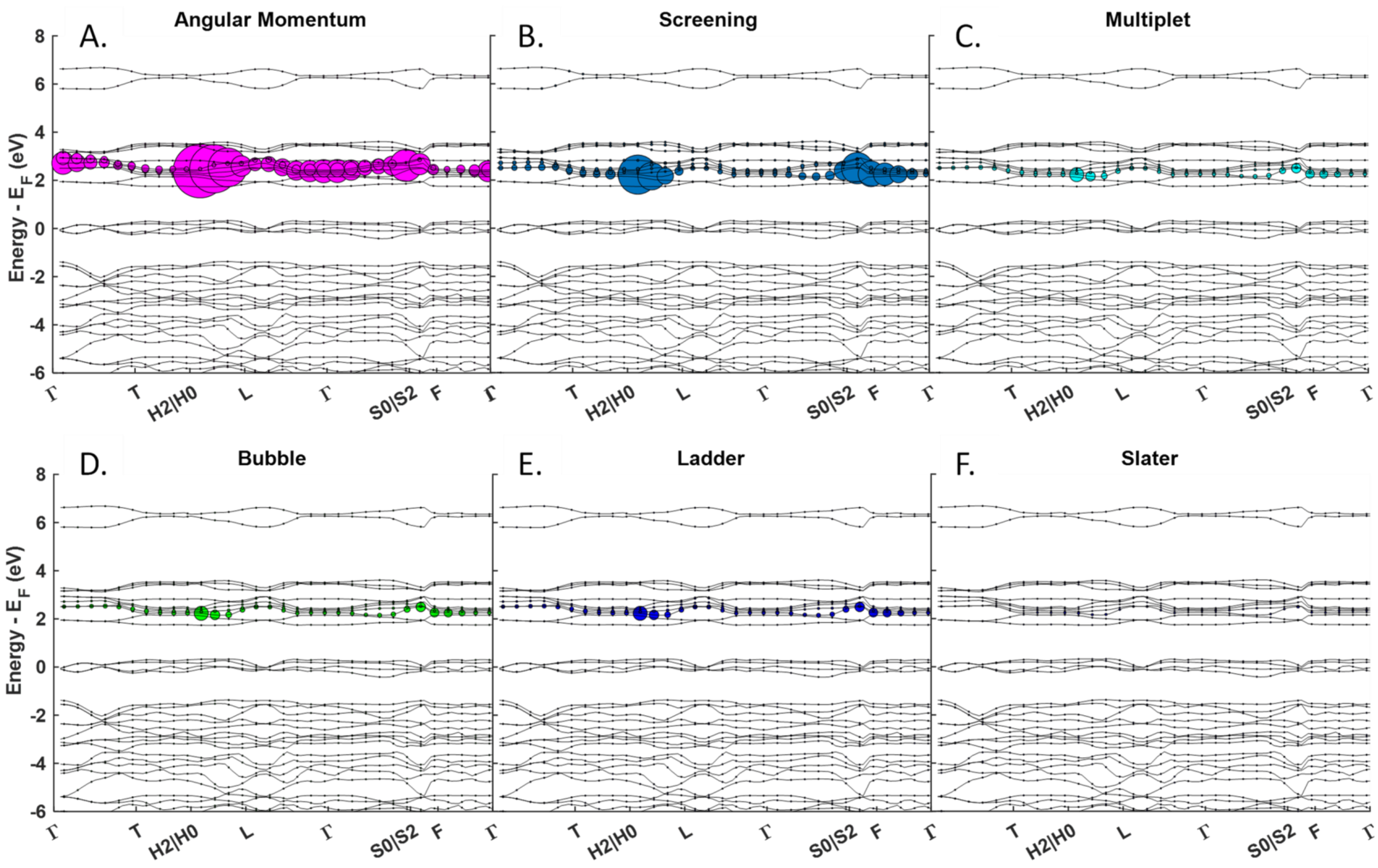

Figure S14. XUV transition Hamiltonian Decomposition of Core-Valence Exciton for the State-Filling Excited State of α-$Fe_2O_3$. The (a) angular momentum, (b) core-hole screening, (c) atomic multiplet, (d) bubble, (e) ladder and (f) Slater core-valence exciton densities are calculated and normalized to determine the relative contribution of each of these processes to the overall XUV exciton and thus absorption spectrum.

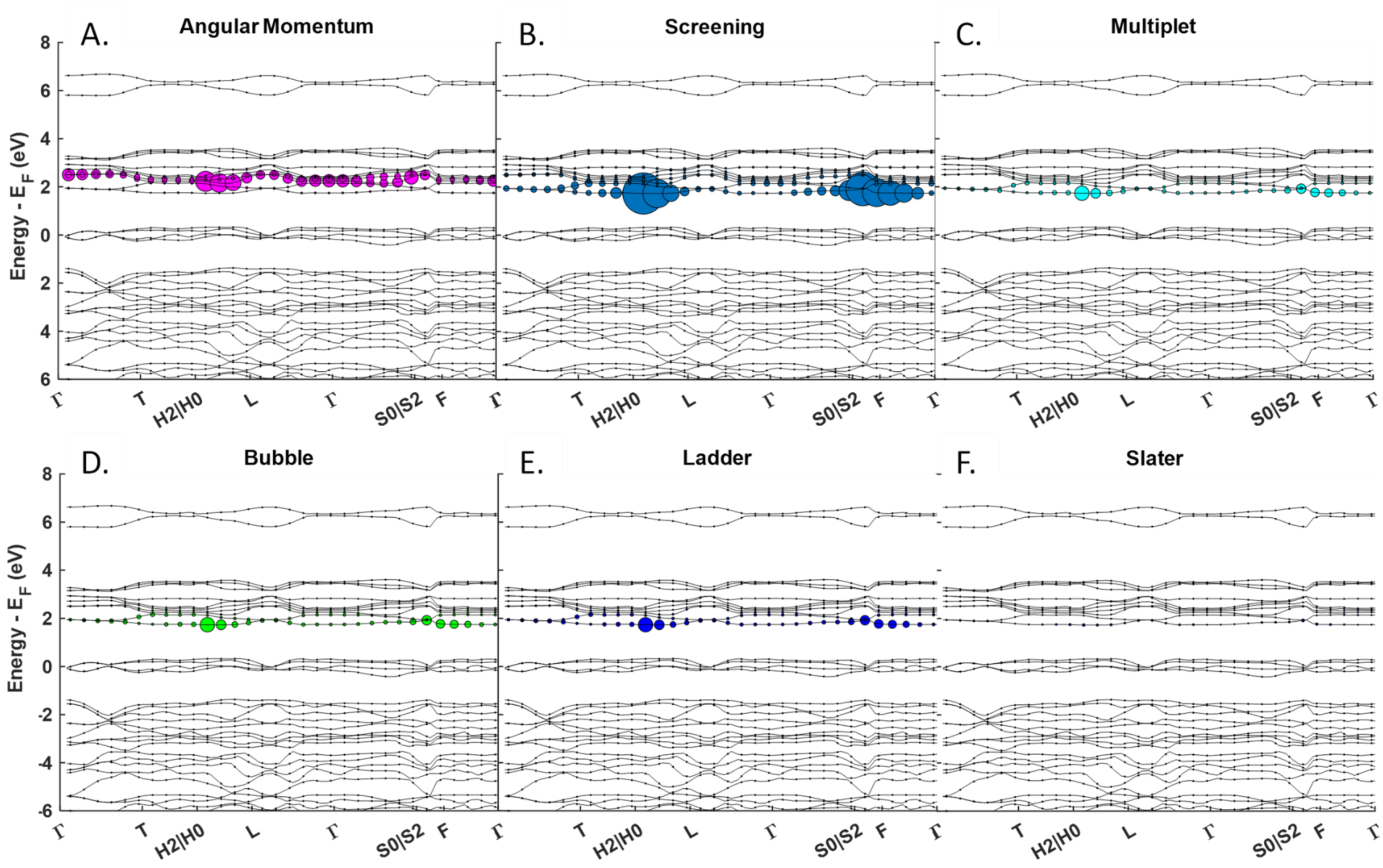

Figure S15. XUV transition Hamiltonian Decomposition of Core-Valence Exciton for the Polaron Excited State of α-$Fe_2O_3$. The (a) angular momentum, (b) core-hole screening, (c) atomic multiplet, (d) bubble, (e) ladder and (f) Slater core-valence exciton densities are calculated and normalized to determine the relative contribution of each of these processes to the overall XUV exciton and thus absorption spectrum.

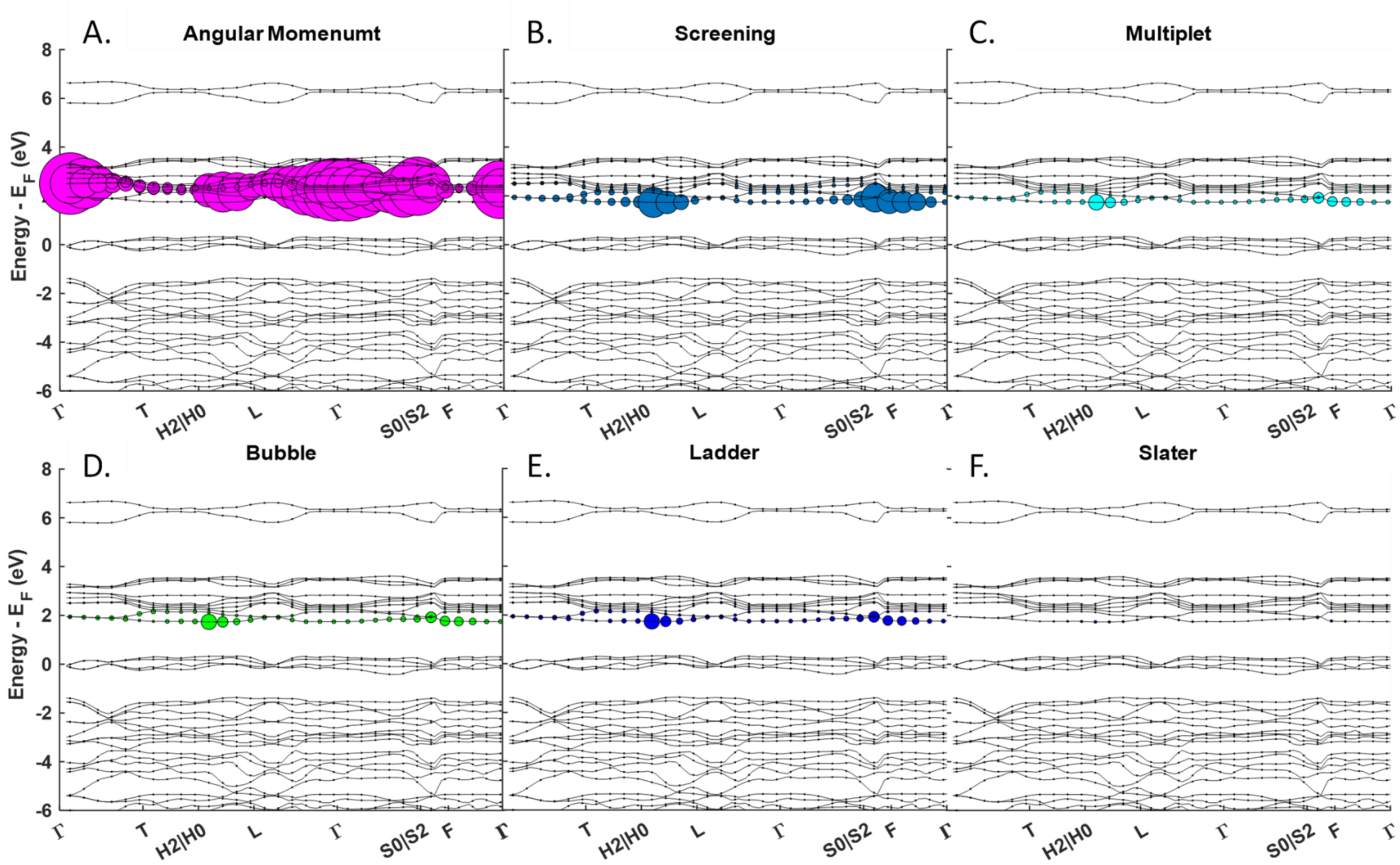

Figure S16. XUV transition Hamiltonian Decomposition of Core-Valence Exciton for the Thermally Expanded Excited State of α-$Fe_2O_3$. The (a) angular momentum, (b) core-hole screening, (c) atomic multiplet, (d) bubble, (e) ladder and (f) Slater core-valence exciton densities are calculated and normalized to determine the relative contribution of each of these processes to the overall XUV exciton and thus absorption spectrum.

c. Comparisons of the differential between the total core-valence exciton density for the excited states of $\alpha$-$Fe_2O_3$ and the ground state modeled in this study.

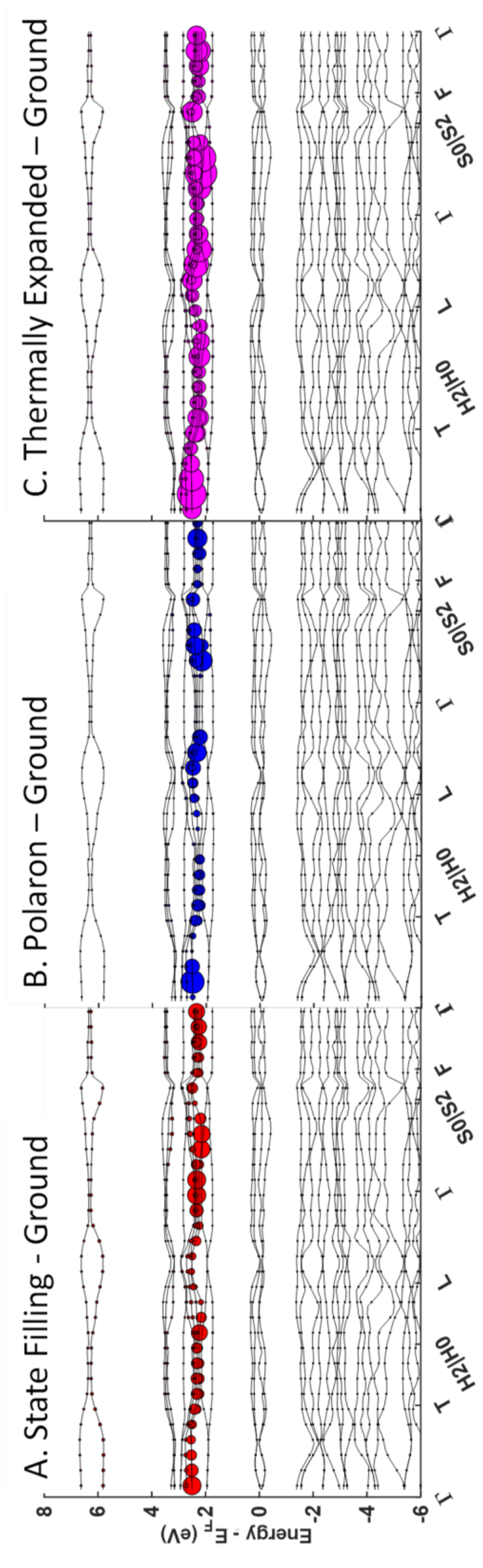

Figure S17. Differential Total XUV transition Density for $\alpha$-$Fe_2O_3$ Excited States. Differentials for the (a) state filling, (b) polaron, and (c) thermally expanded excited states.

d. Differential Hamiltonian decompositions for the state filling, polaron, and thermally expanded excited states relative to the ground state are used to quantitatively determine the importance of different components in the transient changes observed experimentally. All contributions are normalized relative to each other to facilitate direct, quantitative comparison.

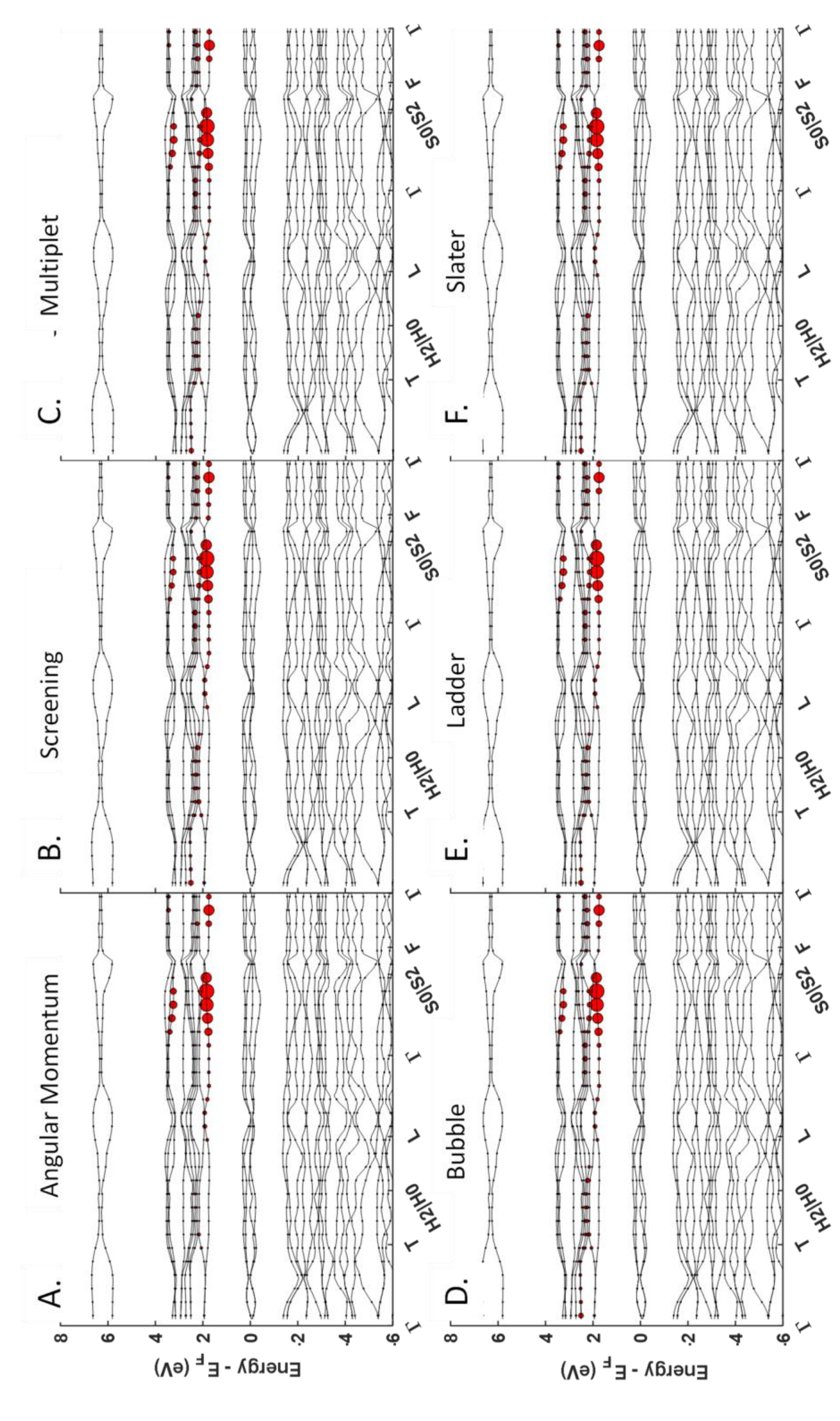

Figure S18. Differential XUV transition Hamiltonian Decomposition of Core-Valence Exciton for the State-Filling Excited State of $\alpha$-$Fe_2O_3$. The magnitude of the state-filling minus the ground state components for the (a) angular momentum, (b) core-hole screening, (c) atomic multiplet, (d) bubble, (e) ladder and (f) Slater core-valence exciton densities are calculated and normalized to determine the relative contribution of each of these processes to excited state effects

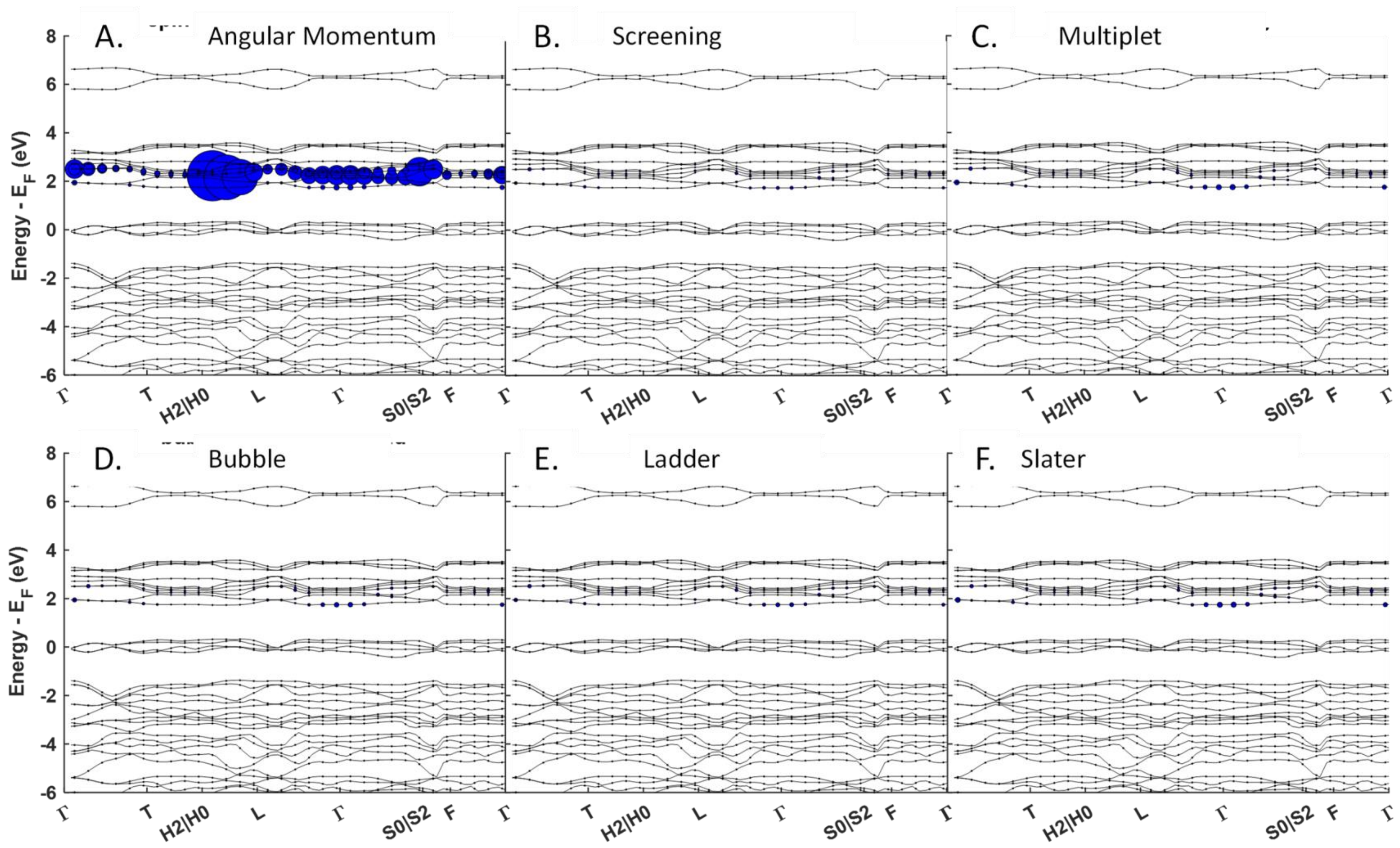

Figure S19. Differential XUV transition Hamiltonian Decomposition of Core-Valence Exciton for the Polaron Excited State of α-$Fe_2O_3$. The magnitude of the polaron minus the ground state for the components of the (a) angular momentum, (b) core-hole screening, (c) atomic multiplet, (d) bubble, (e) ladder and (f) Slater core-valence exciton densities are calculated and normalized to determine the relative contribution of each of these processes to excited state effects.

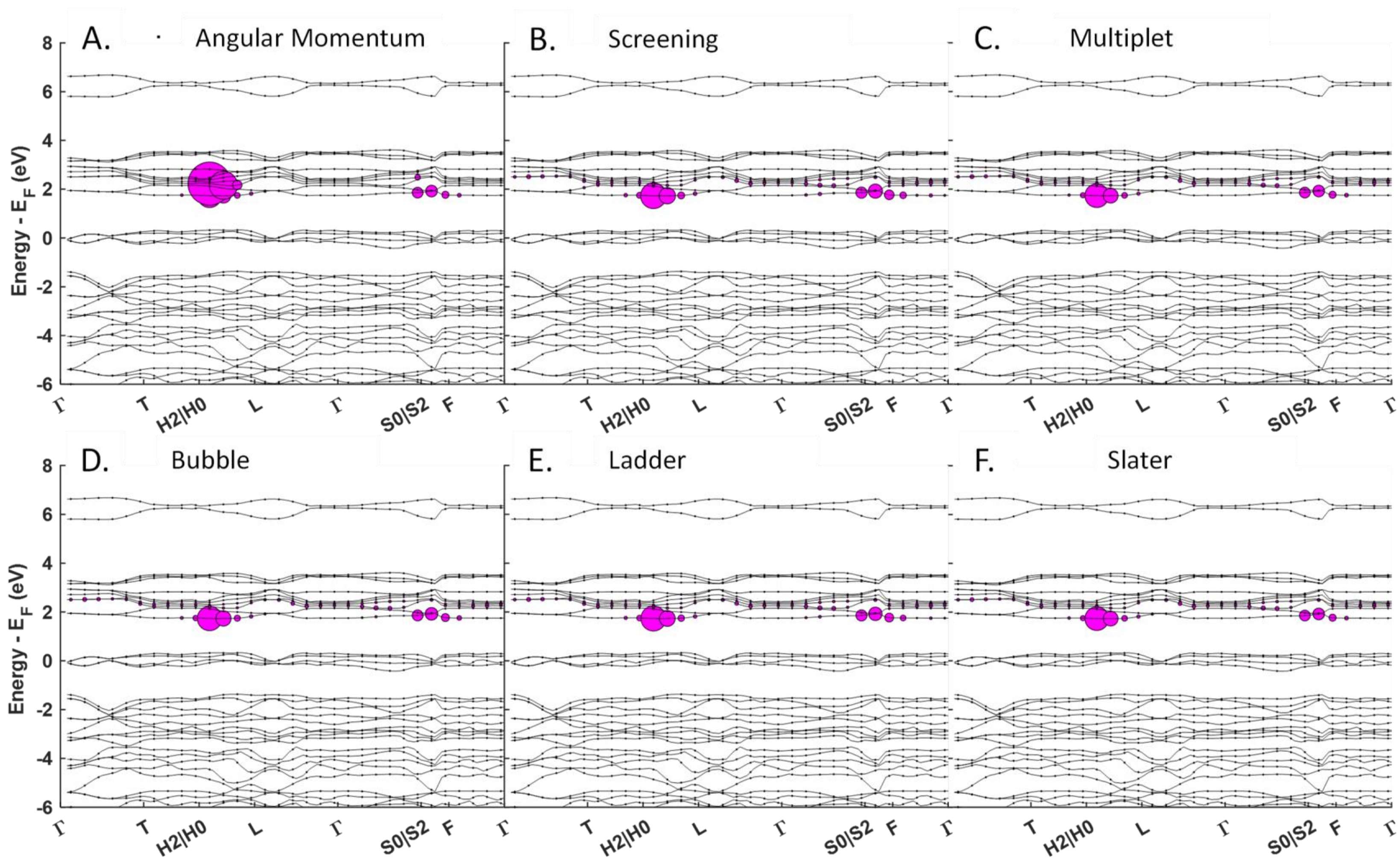

Figure S20. Differential XUV transition Hamiltonian Decomposition of Core-Valence Exciton for the Thermally Expanded Excited State of $\alpha$-$Fe_2O_3$. The magnitude of the thermally expanded minus the ground state for the components of the (a) angular momentum, (b) core-hole screening, (c) atomic multiplet, (d) bubble, (e) ladder and (f) Slater core-valence exciton densities are calculated and normalized to determine the relative contribution of each of these processes to excited state effects.

## 5. Excited State Spectra

a. Excited state spectra were calculated as described in detail in the text, using the same computational parameters as the ground state calculation. Excited state broadening was performed consistent with previously described ground state broadening in SI section 3b.

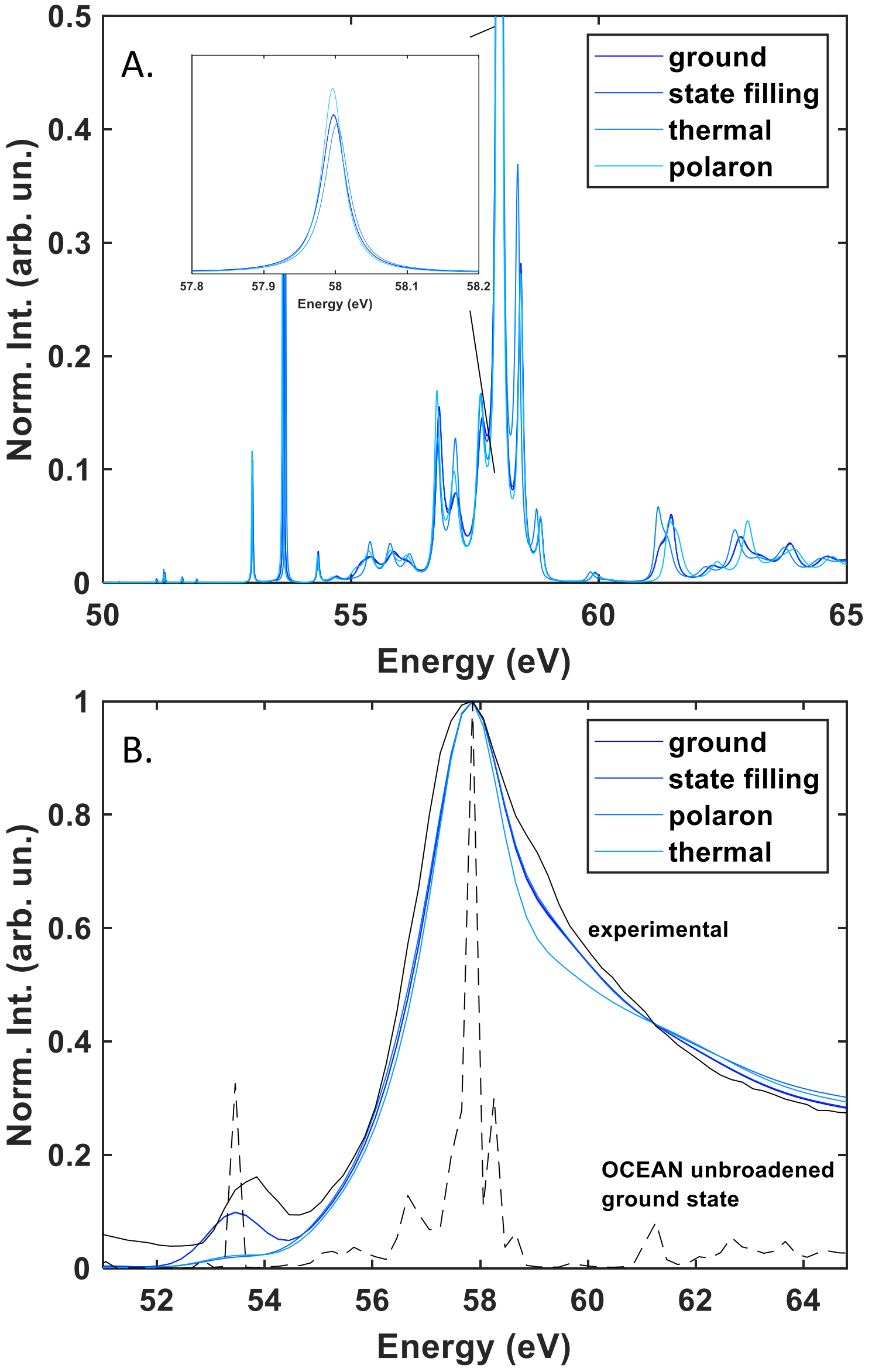

Figure S21. Calculated Spectra for Ground, State Filling, Polaron, and Thermally Expanded States of α-$Fe_2O_3$. (a) Unbroadened OCEAN output XUV absorption spectra for these states. Inset zooms in on the main peak showing how it shifts in energy. (b) Broadened OCEAN outputs compared to the experimental ground state.

b. A simple exponential broadening scheme to model the differential curve between the excited state and the ground state spectra was shown to produce relatively good agreement between the calculated and experimental data. In the exponential broadening model, the calculated XUV absorption spectra for the ground state and state filling (or polaron) are subtracted (without any initial broadening), and the difference is broadened with an exponential. While this broadening scheme clearly over broadens the differential, it also clearly reproduces the signatures of these two states.

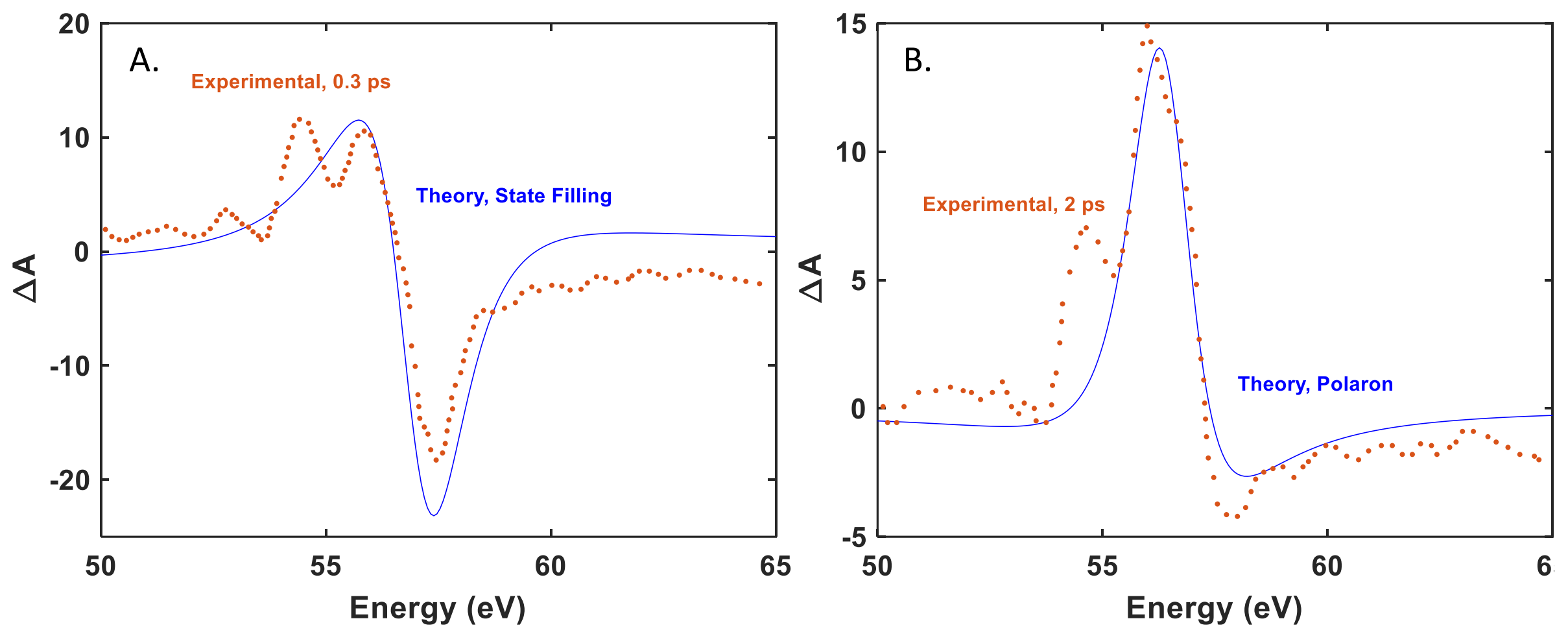

Figure S22. Calculated Differential Spectra for Excited States of $\alpha$-$Fe_2O_3$. (a) Comparison between experimental differential after 0.3 ps and state filling differential for exponential broadening. (b) Comparison between experimental differential after 2 ps and polaron differential for exponential broadening.

c. Broadened differential traces, compared without normalization to demonstrate the relative intensities of the different differential signals.

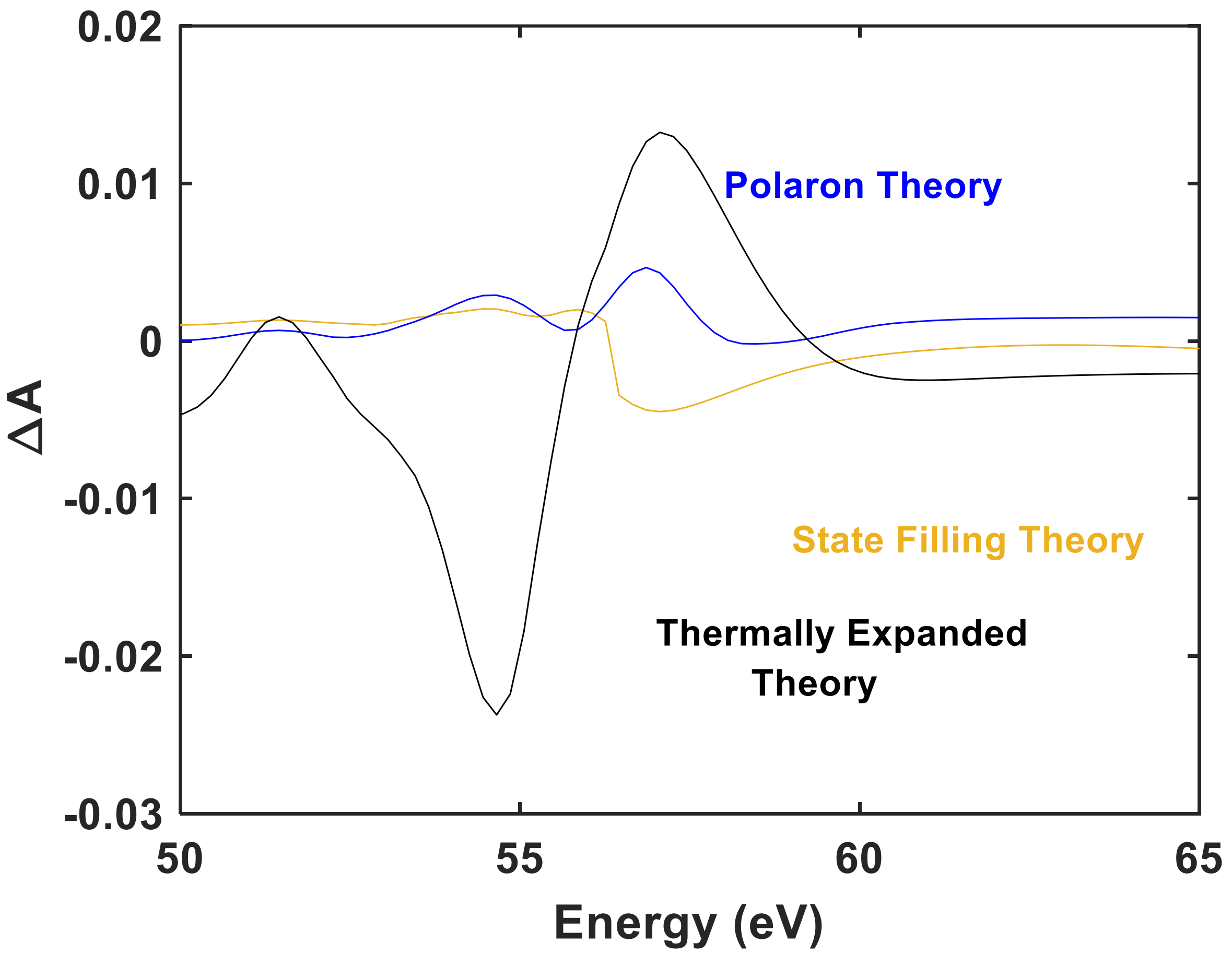

Figure S23. Differential Signals for state filling, polaron, and thermally expanded models of the excited states of $\alpha-Fe_2O_3$ without normalization.

d. Comparison between calculated polaron states with and without additional state filling shows the relative dominance of the polaron in determining transient signals.

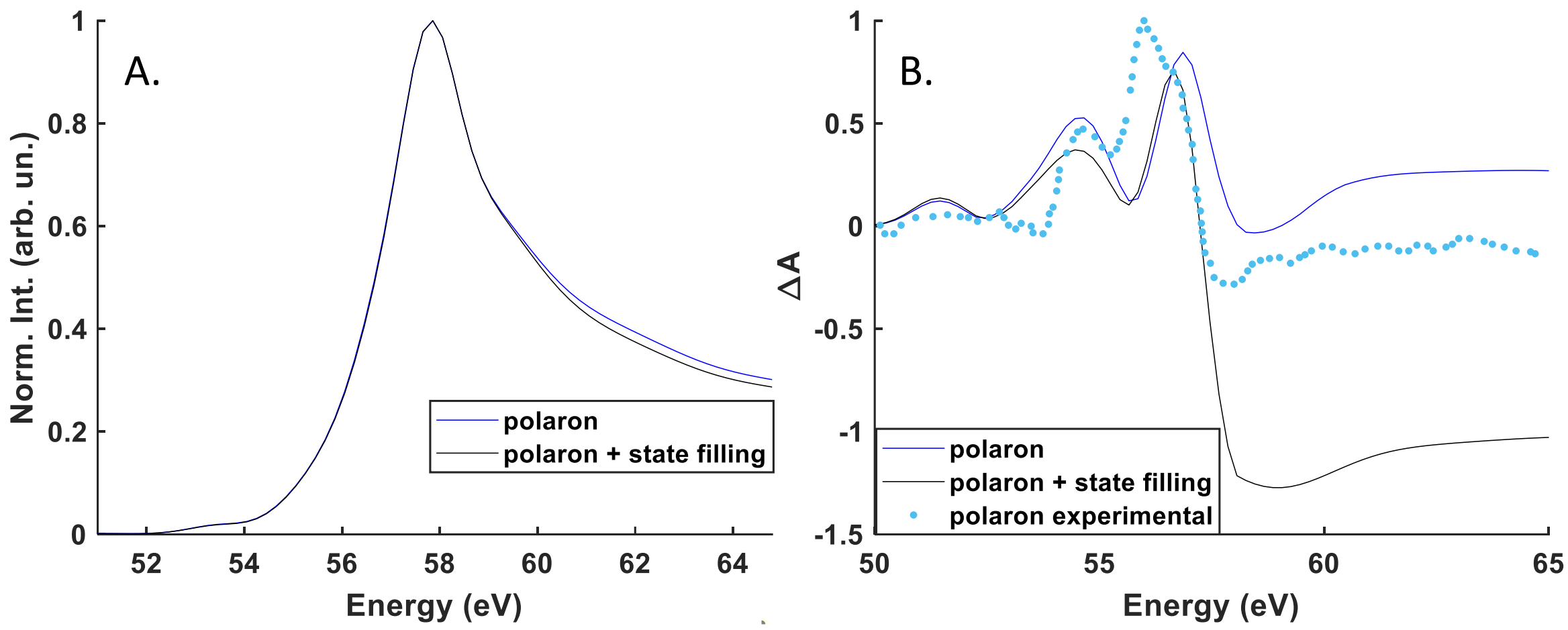

Figure S24. Polaron States of α-$Fe_2O_3$ with and without State Filling. (a) Initial calculated absorption with energy-dependent Gaussian broadening and (b) differential with additional energy-dependent amplitude corrections. The differential plot includes the experimentally measured polaron state differential spectrum.

e. Different configurations were used to model the state filling excited state, using excitation from 0 eV to 1.8 eV above the conduction band minimum. The electron and hole occupations resulting from excitation with different energies were used in the OCEAN calculation to generate model state filling XUV absorption spectra.

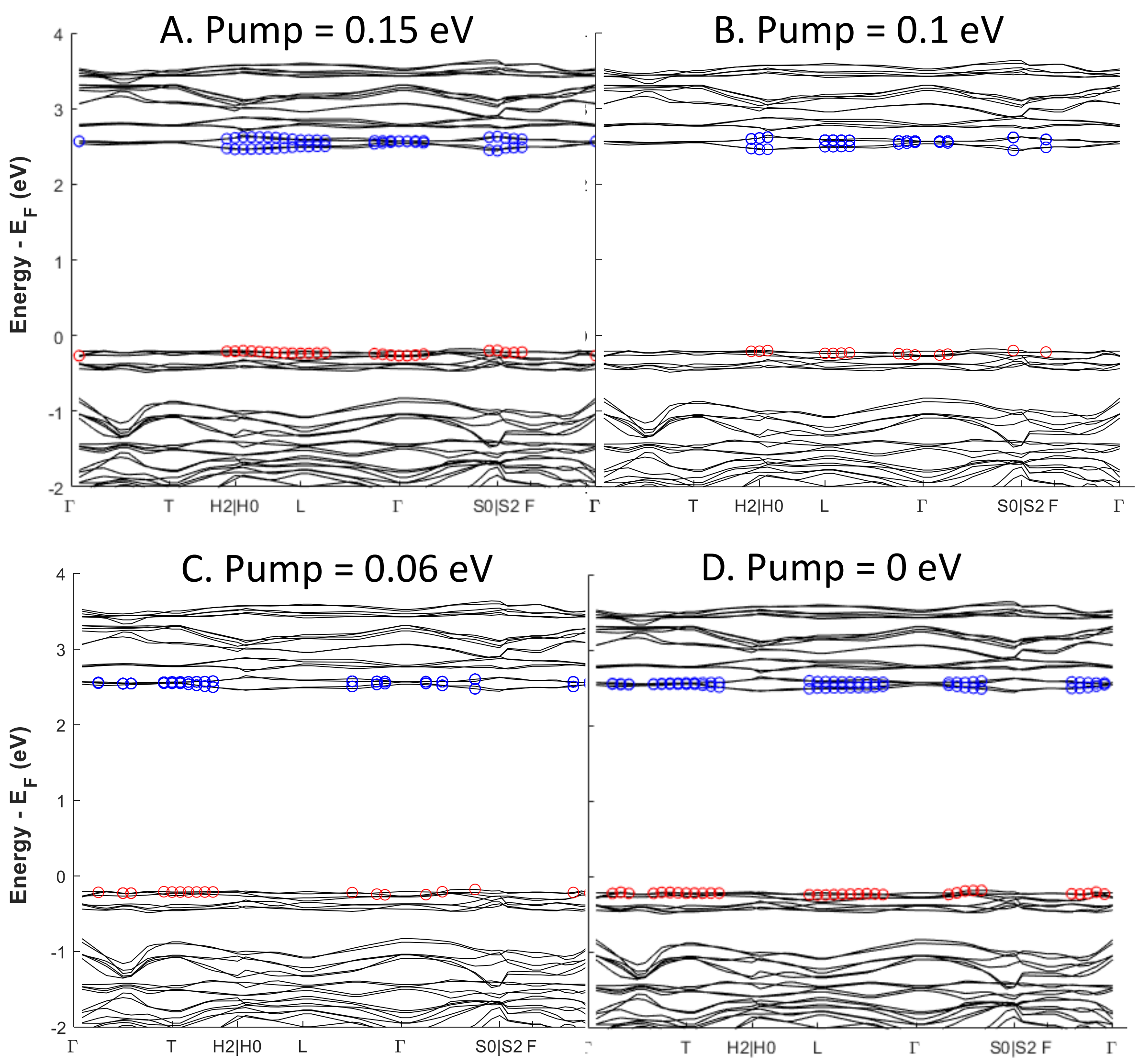

Figure S25. Band Filling for Different Pump Energies Used to Model State Filling. Pump energies from (a) 0.15 eV, (b) 0.1 eV, (c) 0.06 eV, and (d) 0 eV above the conduction band minimum were used to model state filling, as shown in Figure 4. Blue circles indicate electron occupation and red circles indicate holes.

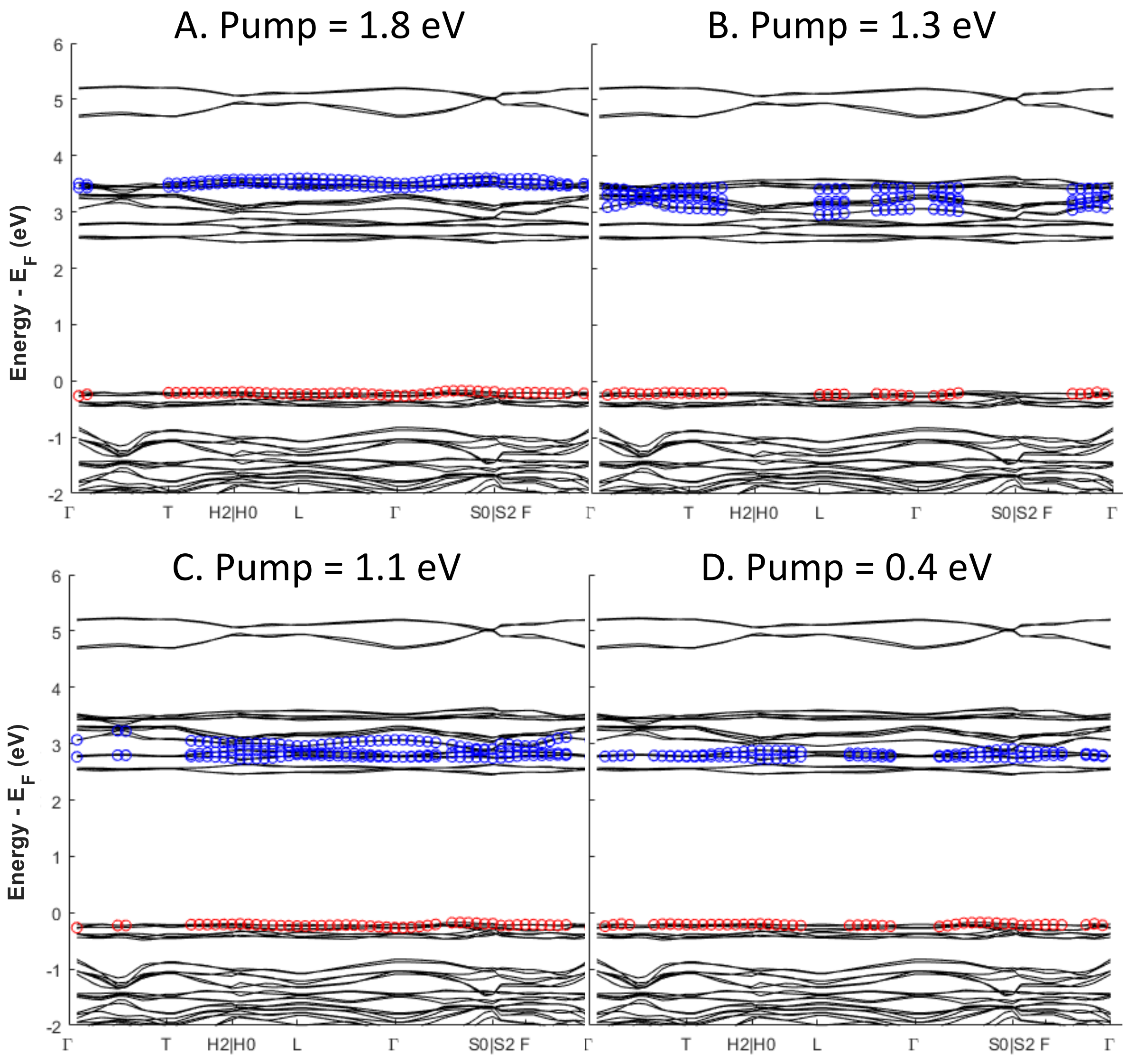

Figure S26. Band Filling for Different Pump Energies Used to Model State Filling. Pump energies from (a) 1.8 eV, (b) 1.3 eV, (c) 1.1 eV, and (d) 0.4 eV above the conduction band minimum were used to model state filling, as shown in Figure S28. Blue circles indicate electron occupation and red circles indicate holes.

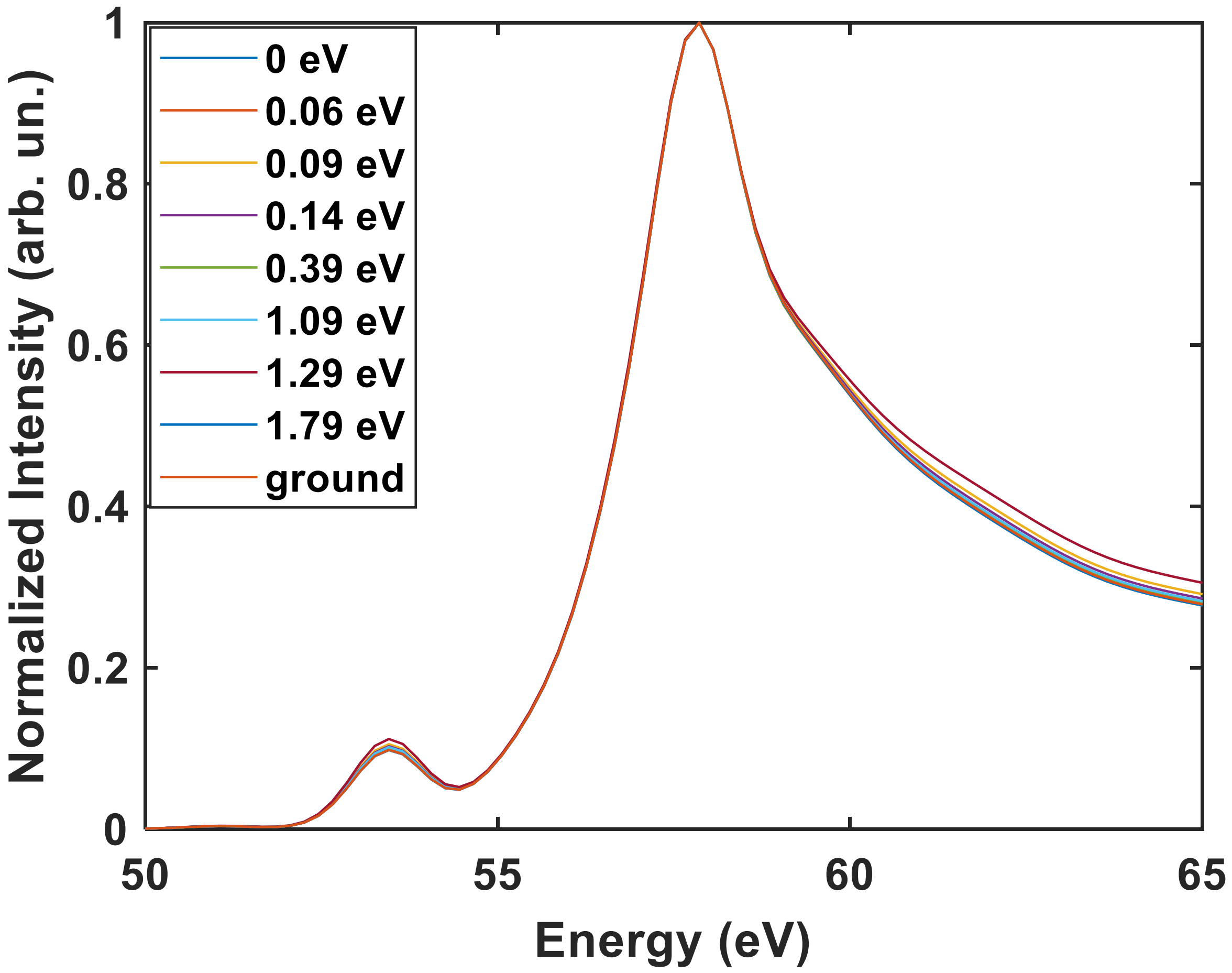

Figure S27. Excitation Energy Dependent XUV Absorption Calculations. Spectra at the Fe $M_{2,3}$ edge of $\alpha$-$Fe_2O_3$ following state filling with different excitation energies. Energy above the conduction band minimum for each trace is shown in eV in the legend.

## 6. References

(1) Attar, A. R.; Chang, H.-T.; Britz, A.; Zhang, X.; Lin, M.-F.; Krishnamoorthy, A.; Linker, T.; Fritz, D.; Neumark, D. M.; Kalia, R. K.; Nakano, A.; Ajayan, P.; Vashishta, P.; Bergmann, U.; Leone, S. R. Simultaneous Observation of Carrier-Specific Redistribution and Coherent Lattice Dynamics in 2H-MoTe$_{2}$ with Femtosecond Core-Level Spectroscopy. *arXiv:2009.00721 [cond-mat]* **2020**.

(2) Piccinin, S. The Band Structure and Optical Absorption of Hematite (α-$Fe_2O_3$): A First-Principles GW-BSE Study. *Phys. Chem. Chem. Phys.* **2019**, *21* (6), 2957–2967. https://doi.org/10.1039/C8CP07132B.

(3) Vura-Weis, J.; Jiang, C.-M.; Liu, C.; Gao, H.; Lucas, J. M.; de Groot, F. M. F.; Yang, P.; Alivisatos, A. P.; Leone, S. R. Femtosecond $M_{2,3}$-Edge Spectroscopy of Transition-Metal Oxides: Photoinduced Oxidation State Change in α-$Fe_2O_3$. *J. Phys. Chem. Lett.* **2013**, *4* (21), 3667–3671. https://doi.org/10.1021/jz401997d.

(4) Cushing, S. K.; Zürch, M.; Kraus, P. M.; Carneiro, L. M.; Lee, A.; Chang, H.-T.; Kaplan, C. J.; Leone, S. R. Hot Phonon and Carrier Relaxation in Si(100) Determined by Transient Extreme Ultraviolet Spectroscopy. *Structural Dynamics* **2018**, *5* (5), 054302. https://doi.org/10.1063/1.5038015.